\documentclass[12pt]{elsarticle}

\makeatletter
\def\ps@pprintTitle{%
  \let\@oddhead\@empty \let\@evenhead\@empty
  \def\@oddfoot{\small This manuscript is currently under review for publication in a peer-reviewed journal.}}
\makeatother

\usepackage{amssymb}
\usepackage{amsmath}

\usepackage{lineno}
\usepackage{hyperref}
\usepackage{bm}
\usepackage{upgreek}
\usepackage{multirow}
\usepackage[nameinlink, capitalise]{cleveref}
\usepackage{comment}
\hypersetup{pdfauthor=whatever}

\begin{document}
\sloppy

\begin{frontmatter}

\title{Efficient Bayesian inversion for simultaneous estimation of geometry and spatial field using the Karhunen-Loève expansion}

\author[1]{Tatsuya Shibata\corref{cor1}} %% Author name
\ead{shibata.tatsuya.84z@st.kyoto-u.ac.jp}
\author[1]{Michael C. Koch\fnref{fn1}} %% Author name
\author[2]{Iason Papaioannou} %% Author name
\author[1]{Kazunori Fujisawa} %% Author 

\cortext[cor1]{Corresponding author}

\affiliation[1]{organization={Graduate School of Agriculture, Kyoto University}, %Department and Organization
            addressline={Kitashirakawa-oiwakecho}, 
            city={Sakyo-ku},
            postcode={606-8502 Kyoto}, 
            %state={},
            country={Japan}}
%\affiliation[2]{organization={Present address: Research School of Earth Sciences, Australian National University}, %Department and Organization
            %addressline={142 Mills Rd}, 
            %postcode={0200 Acton ACT}, 
            %country={Australia}}
\affiliation[2]{organization={Engineering Risk Analysis Group, Technische Universität München}, %Department and Organization
            addressline={Arcisstr. 21}, 
            %city={},
            postcode={80290 München}, 
            %state={München},
            country={Germany}}

\fntext[fn1]{Present address: Research School of Earth Sciences, Australian National University, 142 Mills Rd, 0200 Acton ACT, Australia}

%% Abstract
\begin{abstract}
Detection of abrupt spatial changes in physical properties representing unique geometric features such as buried objects, cavities, and fractures is an important problem in geophysics and many engineering disciplines. In this context, simultaneous spatial field and geometry estimation methods that explicitly parameterize the background spatial field and the geometry of the embedded anomalies are of great interest. This paper introduces an advanced inversion procedure for simultaneous estimation using the domain independence property of the Karhunen-Loève (K-L) expansion. Previous methods pursuing this strategy face significant computational challenges. The associated integral eigenvalue problem (IEVP) needs to be solved repeatedly on evolving domains, and the shape derivatives in gradient-based algorithms require costly computations of the Moore-Penrose inverse. Leveraging the domain independence property of the K-L expansion, the proposed method avoids both of these bottlenecks, and the IEVP is solved only once on a fixed bounding domain. Comparative studies demonstrate that our approach yields two orders of magnitude improvement in K-L expansion gradient computation time. Inversion studies on one-dimensional and two-dimensional seepage flow problems highlight the benefits of incorporating geometry parameters along with spatial field parameters. The proposed method captures abrupt changes in hydraulic conductivity with a lower number of parameters and provides accurate estimates of boundary and spatial-field uncertainties, outperforming spatial-field-only estimation methods.
\end{abstract}

\begin{comment}
\begin{highlights}
%\item Spatial field and interface are estimated simultaneously by the Bayesian inversion.
%\item  Karhunen-Loève expansion on the bounding domain makes the inversion efficient.
%\item Numerical experiments show the methodological advantage of this simultaneous estimation procedure.
\item A method for estimating spatial fields and geometry parameters using HMC is shown. 

%\item Repeated IEVP solves are avoided by leveraging the K-L expansion on a bounding domain.
\item The K-L expansion applied on a bounding domain avoids repeated eigenpair evaluations.

\item Interpolation from the bounding domain simplifies gradient computations in HMC.

\item Numerical experiments show the method's benefits compared to previous methods. 

\end{highlights}
\end{comment}

%% Keywords

\end{frontmatter}

%%\linenumbers

%% Section 1
\section{Introduction}
In nondestructive testing procedures, measurements are often used to estimate spatial distributions of physical properties such as hydraulic conductivity \cite{mclaughlin1996, lee2014, yang2019}, elastic modulus \cite{parida2018, koch2020a}, or P- and S-wave velocities \cite{parida2018, fathi2016, tran2013} through the solution of inverse problems. Quite often, along with the general spatial distribution of the physical property, the interest may lie in identifying unique geometric features such as buried objects, cavities, and fractures, etc. These features manifest as anomalies (extremely high or low magnitudes) in the inferred physical properties \cite{putiska2012, cardarelli2006, adamo2021, ahmed2003, hussain2020} and are delineated in the post-processing stage of inversion. We focus on such inverse problems within a Bayesian framework, incorporating priors and computing uncertainties in inversion results from posterior probability distributions. 

\begin{figure}
\centering
\includegraphics[width=1.0\textwidth]{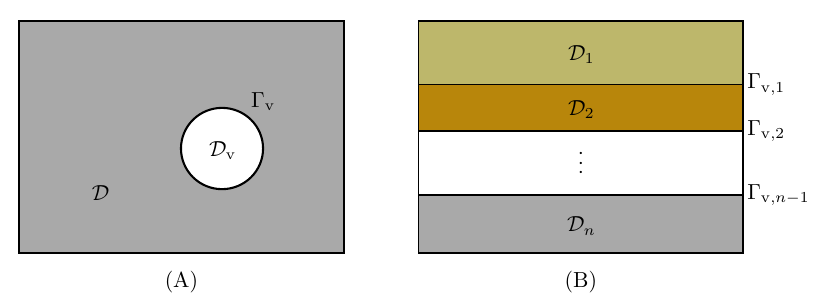}
\caption{\label{fig:two_types_domain}(A) Target area consisting of the soil domain with a circular cavity ($\mathcal{D}' = \mathcal{D} \cup \mathcal{D}_\mathrm{v} $). (B) Target area showing stratification of the subsurface into several domains ($\mathcal{D}' = \bigcup_{i=1}^n \mathcal{D}_{i})$.}
\end{figure}

Consider the inverse problem of visualizing the interior of a target domain with unique geometric features using steady seepage flow observation data. Traditional methods do not account for geometric information, focusing solely on the spatial distribution of hydraulic conductivity. 
However, in practical problems, it is often possible to parametrize the geometries of these features and incorporate a set of geometry parameters into the inverse problem \cite{lahivaara2014estimating,nguyen2018reconstructing,takamatsu2020}.
The general construct of such problems is explained through a specific example involving seepage flow in a domain shown in \cref{fig:two_types_domain}(A). This problem involves the determination of the hydraulic conductivity spatial field in the domain $\mathcal{D}$ around a pipe (of unknown location and size) beneath the ground surface. This inverse problem can help track water leakage leading to subsequent soil erosion in the vicinity of the pipe \cite{sato2015influence}. Instead of determining the spatial field over the entire domain $\mathcal{D}' = \mathcal{D} \cup \mathcal{D}_\mathrm{v}$, it might be beneficial to encode a parameterization of the unknown location and size of the pipe-soil interface $\Gamma_\mathrm{v}$ into the inverse problem and determine the spatial field only over the domain $\mathcal{D}$. Another example shown in \cref{fig:two_types_domain}(B) is a trans-dimensional subsurface planarly-layered stratification problem \cite{malinverno2002parsimonious, cao2019bayesian} where along with the layer properties in each domain $\mathcal{D}_{i}$, $i\in \{1,\ldots,n\}$, geometry parameters defining the planar layer boundaries $\Gamma_{\mathrm{v},i}$, $i\in \{1,\ldots,n-1\}$ can also be determined during inversion. This approach, referred to as the \emph{simultaneous estimation} method, contrasts with \emph{spatial-field-only estimation} methods that neglect explicit geometry parameters.

Incorporating geometry parameters allows for accurate boundary condition enforcement. For instance, in \cref{fig:two_types_domain}(A) the no-flow boundary conditions representative of actual hydraulic conditions on $\Gamma_\mathrm{v}$ can be implemented easily in numerical schemes such as the finite element method. This can be done through finite element meshes that conform to the computational domain $\mathcal{D}$ that is continuously updated as $\Gamma_\mathrm{v}$ is updated during inversion. The challenge in such a shape-tracking approach is maintaining decent mesh quality during successive updates. Drawing on a method developed by Koch et al. (2020) \cite{koch2020b}, we achieve this by updating the domain from a fixed reference configuration using a mesh-moving finite element method \cite{stein2003mesh}. Computation of gradients of the posterior density w.r.t the geometry parameters follows from shape sensitivity analysis \cite{Haslinger2003}.

Detection of sharp interfaces in material properties has historically been solved through level set methods \cite{aghasi2011a, mcmillan2015, iglesias2016} with great success. These methods work on a fixed domain and do not implement boundary conditions (BCs) at the interfaces. However, it must be mentioned that a conformal meshing strategy can easily be applied once the boundary has been defined by the zero contours of a level set following which BCs can be applied. For simplicity and practical utility in problems such as that in \cref{fig:two_types_domain}(A), we restrict ourselves to working with a direct parameterization of the geometry of the boundary. However, the topics discussed in subsequent sections should, in principle, be equally applicable to problems with level-set parameterizations. The explicit consideration of boundaries through these methods can aid the inversion process in detecting sharp interfaces between material domains of vastly varying (or discontinuous) physical properties—a task that is difficult to achieve, especially when the regularization scheme implemented only permits smooth solutions. This is the case in conventional Bayesian inversion techniques where Gaussian random fields with smooth autocovariance functions \cite{williams2006gaussian} are used as priors. In this study, we only consider stationary, isotropic autocovariance functions with a single length scale and note that approaches considering non-stationary autocovariance functions in a hierarchical setting \cite{calvetti2008} are strong alternative approaches to detect interfaces between discontinuous materials.

Our study estimates geometry and spatial field parameters simultaneously, characterizing material properties over the domain $\mathcal{D}$ as a spatial random field and finding its posterior statistics along with those of the geometry parameters defining the boundary $\Gamma_\mathrm{v}$. This is in contrast to studies that aim to determine constant spatial properties once the boundary of an embedded object has been determined through level sets \cite{holbach2023bayesian} or studies that treat background spatial field parameters as ‘nuisance’ parameters \cite{lahivaara2014estimating,belliveau2023}. The prior spatial random field is discretized through the Karhunen-Loève (K-L) expansion \cite{loeve1978, ghanem2012}, which is the optimal orthogonal series representation in the mean-square sense. The eigenvalues and eigenfunctions employed in the expansion are computed by solving an integral eigenvalue problem (IEVP) related to the autocovariance function of the random field defined over $\mathcal{D}$ \cite{betz2014}. 
%It is well known that the eigenvalues of the covariance kernel decay exponentially fast for smooth autocovariance kernels and algebraically fast in other cases \cite{marzouk2009}. 
In practice, the K-L expansion is applied as a truncated series, and the effect of the choice of prior parameters (correlation lengths, number of terms in the truncated expansion, etc.) on the statistics of the posterior has been studied extensively in the literature \cite{uribe2020}.  

In this study, the sampling of the posterior distribution over the random coefficients of the K-L expansion and the geometry parameters (Koch et al. (2021) \cite{koch2021}) is done through the Hamiltonian Monte Carlo (HMC) algorithm \cite{neal2011}. The proposals in this Markov Chain Monte Carlo (MCMC) \cite{metropolis1953equation,hastings1970monte,gamerman2006markov} algorithm utilize gradient information of the target distribution and can be tuned \cite{hoffman2014,betancourt2018} to draw nearly independent samples from the posterior. This is especially useful in computationally expensive PDE-constrained inverse problems where random walks must be avoided.
To satisfy the reversibility condition \cite{chib1995understanding} of MCMC, a non-trivial task for simultaneous spatial field and geometry updates, we employ mesh-moving updates \cite{koch2020b} of the finite element mesh from a fixed reference configuration. Finally, updating the boundary $\Gamma_\mathrm{v}$ requires redefining the autocovariance function and computing the IEVP over the
updated domain $\mathcal{D}$, leading to a high computational cost as the eigenpair related to the IEVP needs to be recomputed. This problem is exacerbated in HMC, where the gradients of the posterior density require a computation of the gradients of the eigenpair from the IEVP w.r.t the geometry parameters \cite{koch2021}. This is particularly demanding because the evaluation of the shape derivatives requires the computation of the inverse of a rank-deficient matrix related to the eigenvalue problem, thereby necessitating the computation of pseudo-inverses (e.g. Moore-Penrose inverse \cite{magnus1985}).

In this study, we revisit the simultaneous estimation method of Koch et al. (2021) \cite{koch2021} with the aim of improving computational efficiency. Leveraging the domain independence property of the K-L expansion \cite{pranesh2015}, which states that the first and second-order moments of a random field generated by the K-L expansion are invariant to a change in the physical domain, this study aims to:
\begin{itemize}
    \item Develop a Bayesian inversion procedure for simultaneous spatial field and geometry inversion using conformal meshes. This includes the development of a novel gradient computation procedure. 
    \item Investigate the difference in inversion results and computation time between the simultaneous estimation method of Koch et al. (2021) \cite{koch2021} and the method developed in this study.
    \item Highlight the advantage of explicitly considering a parameterization of the geometry in posterior statistics as compared to general spatial-field-only estimation methods.
\end{itemize}

The subsequent sections of this study are structured as follows. \cref{sec:SpatialField} explains the parameterization of the targets to be estimated, including the K-L expansion and its domain independence property. In \cref{sec:Bayes}, the forward and observation models are first described. Following this, the framework for Bayesian inference including HMC, is explained. \cref{sec:grad} explains the gradient computation procedure. In \cref{sec:NumExp}, numerical analysis results of inverse analysis for one-dimensional and two-dimensional seepage flow problems are presented.

%% ==================================================================
%% =========================== Section 2 ============================
%% ==================================================================
\section{Modelling of the spatial random field and geometry}\label{sec:SpatialField}

%% ==================================================================
\subsection{Karhunen-Loève expansion}\label{subsec:KL}
Let $\left(\Omega,\mathcal{\mathcal{F}},P\right)$ be a probability space where $\Omega$ is the sample space, $\mathcal{\mathcal{F}}$
is the $\sigma$ algebra over $\Omega$, and $P$ is a probability
measure on $\mathcal{\mathcal{F}}$. Also, let $\mathcal{D\subset}\mathbb{R}^{d}$
be a bounded domain. Consider a real-valued, second-order spatial random field $X\left(\mathbf{z},\omega\right):\mathcal{D}\times\Omega\rightarrow\mathbb{R}$ with a continuous mean $\overline{X}\left(\mathbf{z}\right):\mathcal{D\rightarrow\mathbb{R}}$
and a continuous autocovariance function $C\left(\mathbf{z},\mathbf{z}^{*}\right):\mathcal{D}\times\mathcal{D}\rightarrow\mathbb{R}$. The autocovariance function is a positive-semidefinite function which, according to Mercer's theorem, has the spectral decomposition \cite{ghanem2012}:
\begin{equation}
C\left(\mathbf{z},\mathbf{z}^{*}\right)=\sum_{i=1}^{\infty}\lambda_{i}\phi_{i}\left(\mathbf{z}\right)\phi_{i}\left(\mathbf{z}^{*}\right),\label{eq:Mercer}
\end{equation}
where $\lambda_{i}$ and $\phi_{i}$ are eigenvalues and eigenfunctions
of the Hilbert-Schmidt integral operator corresponding to $C$, and
the eigenfunctions are orthogonal, that is, $\int_{\mathcal{D}}\phi_{i}(\mathbf{z})\phi_{j}(\mathbf{z})\mathrm{d}\mathbf{z}=\delta_{ij}$.
The eigenpairs $\{\lambda_{i},\phi_{i}\}_{i=1}^\infty$ are computed as solutions to the following integral eigenvalue problem (IEVP), which is a homogeneous Fredholm integral equation of the second kind \cite{le2010spectral}:
\begin{equation}
\int_{\mathcal{D}}C\left(\mathbf{z},\mathbf{z}^{*}\right)\phi_{i}(\mathbf{z}^{*})\mathrm{d}\mathbf{z}^{*}=\lambda_{i}\phi_{i}(\mathbf{z}).\label{eq:IEVP}
\end{equation}

With these results, it can be shown that the second-order random field $X\left(\mathbf{z},\omega\right)$ can be represented through the K-L expansion as follows:
\begin{equation}
X\left(\mathbf{z},\omega\right)=\Xi\left(\mathbf{z},\omega\right)\equiv\overline{X}\left(\mathbf{z}\right)+\sum_{i=1}^{\infty}\sqrt{\lambda_{i}}\phi_{i}\left(\mathbf{z}\right){}^{1}\theta_{i}\left(\omega\right),\label{eq:KL}
\end{equation}
where $^{1}\theta_{i}\left(\omega\right)=\frac{1}{\sqrt{\lambda_{i}}}\int_{\mathcal{D}}\left(X\left(\mathbf{z},\omega\right)-\overline{X}\left(\mathbf{z}\right)\right)\phi_{i}\left(\mathbf{z}\right)\mathrm{d}\mathbf{z}$
are random coefficients that have the following properties: $^{1}\theta_{i}$ satisfies $E\left[^{1}\theta_{i}\right]=0,E\left[^{1}\theta_{i}{}^{1}\theta_{j}\right]=\delta_{ij}$.
The work in this paper applies the K-L expansion to Gaussian random fields $X\left(\mathbf{z},\omega\right)$. In this case, since the random variables $^{1}\theta_{i}\left(\omega\right)$ are linear functions of the Gaussian field, they are independent standard normal random variables. 
%,i.e., $^{1}\theta_{i}\sim\mathcal{N}(0,1)$. 
In \cref{eq:KL}, $\Xi\left(\mathbf{z},\omega\right)$ is
the complete K-L expansion of $X\left(\mathbf{z},\omega\right)$
and is defined to distinguish it from the truncated K-L expansion
$\hat{\Xi}\left(\mathbf{z},\omega\right)$. If $\left\{ \lambda_{i}\right\} _{i=1}^{\infty}$ are sorted in descending
order, $\lambda_{i}>\lambda_{i+1}$ ($\lim_{i\rightarrow\infty}\lambda_{i}=0$),
the truncated K-L expansion of $X\left(\mathbf{z},\omega\right)$
is represented as
\begin{equation}
X\left(\mathbf{z},\omega\right)\approx\hat{\Xi}\left(\mathbf{z},\omega\right)\equiv\overline{X}\left(\mathbf{z},\omega\right)+\sum_{i=1}^{M_1}\sqrt{\lambda_{i}}\phi_{i}\left(\mathbf{z}\right){}^{1}\theta_{i}\left(\omega\right),\label{eq:truncated_KL}
\end{equation}
where $M_1$ is the number of terms considered in the expansion and
should be appropriately chosen according to the covariance structure of the random field. 
This finite series representation of the random process is optimal in the sense of minimizing the mean-square error \cite{ghanem2012}. In Bayesian inversion, inference over the spatial random field $X\left(\mathbf{z},\omega\right)$, discretized in terms of the truncated K-L expansion, is then equivalent to inference over the apriori standard multivariate Gaussian random variable 
$^{1}\bm{\uptheta} = \left[ \theta_1, \ldots, \theta_{M_1} \right]^\top$.

%% ==================================================================
\subsection{Domain independence property}\label{subsec:DomainIndep}

Let $\mathcal{D}'$ be a domain bounding $\mathcal{D}$  (i.e., $\mathcal{D}'\supset \mathcal{D}$) and $X'\left(\mathbf{z},\omega\right):\mathcal{D}'\times\Omega\rightarrow\mathbb{R}$ be a real-valued and second-order random process with the same mean
and covariance function as $X\left(\mathbf{z},\omega\right)$.
Also, let $\Xi'\left(\mathbf{z},\omega\right)$ denote the K-L expansion of $X'\left(\mathbf{z},\omega\right)$, that is,
\begin{equation}
X'\left(\mathbf{z},\omega\right)=\Xi'\left(\mathbf{z},\omega\right)\equiv\overline{X}\left(\mathbf{z}\right)+\sum_{i=1}^{\infty}\sqrt{\lambda'_{i}}\phi'_{i}\left(\mathbf{z}\right){}^{1}\theta_{i}\left(\omega\right),\label{eq:KL_D'}
\end{equation}
where $\lambda'_{i}$ and $\phi'_{i}$ are eigenvalues and eigenfunctions of IEVP \cref{eq:IEVP} on $\mathcal{D}'$. Since the autocovariance function $C\left(\mathbf{z},\mathbf{z}^{*}\right)$ corresponding to locations common to two domains, i.e., $\mathbf{z},\mathbf{z}^{*}\in \mathcal{D}(= \mathcal{D}\cap\mathcal{D}')$, is the same, Mercer's theorem can be represented as
\begin{equation}
C\left(\mathbf{z},\mathbf{z}^{*}\right)=\sum_{i=1}^{\infty}\lambda_{i}\phi_{i}\left(\mathbf{z}\right)\phi_{i}\left(\mathbf{z}^{*}\right)=\sum_{i=1}^{\infty}\lambda'_{i}\phi'_{i}\left(\mathbf{z}\right)\phi'_{i}\left(\mathbf{z}^{*}\right),\quad \forall \mathbf{z},\mathbf{z}^{*}\subset \mathcal{D}(=\mathcal{D}\cap\mathcal{D}').\label{eq:Mercer2}
\end{equation}
Then from \cref{eq:KL,eq:KL_D',eq:Mercer2}, for the two K-L expansions $\Xi\left(\mathbf{z},\omega\right)$ and $\Xi'\left(\mathbf{z},\omega\right)$, the following equations hold for all $\mathbf{z},\mathbf{z}^{*}\in \mathcal{D}$:
%\begin{equation}
%E\left[\Xi\left(\mathbf{z},\omega\right)\right]=E\left[\Xi'\left(\mathbf{z},\omega\right)\right],
%\end{equation}
%\begin{equation}
%E\left[\Xi^{2}\left(\mathbf{z},\omega\right)\right]=E\left[\Xi'^{2}\left(\mathbf{z},\omega\right)\right],
%\end{equation}
%\begin{equation}
%E\left[\Xi\left(\mathbf{z},\omega\right)\Xi\left(\mathbf{z}^{*},\omega\right)\right]=E\left[\Xi'\left(\mathbf{z},\omega\right)\Xi'\left(\mathbf{z}^{*},\omega\right)\right].
%\end{equation}
\begin{align}
E\left[\Xi\left(\mathbf{z},\omega\right)\right]&=E\left[\Xi'\left(\mathbf{z},\omega\right)\right],\\
E\left[\Xi^{2}\left(\mathbf{z},\omega\right)\right]&=E\left[\Xi'^{2}\left(\mathbf{z},\omega\right)\right],\\
E\left[\Xi\left(\mathbf{z},\omega\right)\Xi\left(\mathbf{z}^{*},\omega\right)\right]&=E\left[\Xi'\left(\mathbf{z},\omega\right)\Xi'\left(\mathbf{z}^{*},\omega\right)\right].  
\end{align}
This is called the domain independence property \cite{pranesh2015} of the K-L expansion, which states that the first and second-order moments of two random fields, generated on two overlapping domains $\mathcal{D}$ and $\mathcal{D}'$ by K-L expansions, are the same on $\mathcal{D} \cap \mathcal{D}'$. In other words, even though the individual K-L terms might differ, the first and second-order moments of the random fields (or the Gaussian random field itself) generated by the K-L expansions are invariant to a change in the physical domain. Note that the domain independence property holds for the complete K-L expansion, and an error \cite{pranesh2015} is introduced when the truncated K-L expansions
$\hat{\Xi}\left(\mathbf{z},\omega\right)$ and $\hat{\Xi'}\left(\mathbf{z},\omega\right)$ are used. 
%The domain independence property is useful from the viewpoint of the computational cost for simultaneously estimating both the physical properties represented as K-L expansion and the geometry using HMC.
The behavior of this error for both the K-L expansions has been studied extensively \cite{papaioannou2012non}. Numerical tests on several examples have shown that the truncation error is larger for the K-L expansion of the random field defined on the bounding domain. This implies that the K-L expansion loses its optimality for representing the random field on $\mathcal{D}$ when the K-L expansion is performed on $\mathcal{D}'$.

%% ==================================================================
\subsection{Parameterization} \label{subsec:parameterization}
Let ${u}\left(\mathbf{z},\omega\right):\mathcal{D}\times\Omega\rightarrow\mathbb{R}$ be a Gaussian random field, representing a spatially varying physical property, 
with a known mean and autocovariance function. Then, according to the K-L expansion in \cref{eq:truncated_KL}, this Gaussian random field is implicitly represented through the parameter vector $^{1}\bm{\uptheta}\in\mathbb{R}^{M_1}$ and endowed with a Gaussian prior density $\mathcal{N}\left(\mathbf{0},\mathbf{I}_{M_1}\right)$. 
Additionally, let $^{2}\bm{\uptheta} \in\mathbb{R}^{M_2}$ be the parameters chosen to define the geometric features of interest.
The choice of parameters depends on the prior information available about the complexity of features in question. This can include direct definitions of simple shapes \cite{koch2020b} or more flexible parameterizations controlling the degree of continuity, slope, and curvature of the boundary such as B-splines \cite{bugeda1993general}.
The distribution of $^{2}\bm{\uptheta}$ must be constrained to a physically sensible range. In this study, the multivariate truncated normal distribution is chosen as the prior of $^{2}\bm{\uptheta}$.
If $\bm{\uptheta} \in \mathbb{R}^M$ is the vector of all the parameters to be estimated, then in subsequent sections, $\bm{\uptheta}={}^{1}\bm{\uptheta}$ in the spatial-field-only estimation method, and $\bm{\uptheta}=\left[^{1}\bm{\uptheta},{}^{2}\bm{\uptheta}\right]^{\top}$ in the simultaneous estimation method. 
The inverse analysis is performed by determining the posterior distribution of the multivariate random variable $\bm{\uptheta}$.

For the simultaneous estimation method, it is easy to see that the computational domain is a function of the geometry parameters, i.e., $\mathcal{D}\left({}^{2}\bm{\uptheta}\right)$. A naive implementation of the K-L expansion to discretize ${u}\left(\mathbf{z},\omega\right)$ on domain configurations updated during inversion would require repeated evaluations of the IEVP in \cref{eq:IEVP}. Additionally, gradient-based Bayesian inversion algorithms such as HMC would require the expensive computation of a Moore-Penrose inverse to obtain the gradient w.r.t the geometry parameters (see \cref{eq:derivative_discreteKL} in \ref{app:discreteKL}).
The domain independence property offers an alternative strategy where it is not necessary to evaluate the IEVP and its derivatives for each realization $\mathcal{D}\left({}^{2}\bm{\uptheta}\right)$. A proper choice of another domain $\mathcal{D}' \supset \mathcal{D}\left({}^{2}\bm{\uptheta}\right)$, such as a bounding box (without holes), that bounds all possible realizations of $\mathcal{D}\left({}^{2}\bm{\uptheta}\right)$ as the geometry parameters are updated, has to be made.
As all coordinates $\mathbf{z}$ in $\mathcal{D}\left({}^{2}\bm{\uptheta}\right)$ also lie in $\mathcal{D}'$, then due to the domain independence property, the K-L expansion of ${u}\left(\mathbf{z},\omega\right)$ constructed on $\mathcal{D}'$ is equivalent to that on $\mathcal{D}\left({}^{2}\bm{\uptheta}\right)$. The eigenpairs are obtained by solving the IEVP only once on the bounding domain at the beginning of inversion, and the eigenfunctions can then be interpolated to points of interest in $\mathcal{D}\left({}^{2}\bm{\uptheta}\right)$. This strategy simplifies the gradient computation procedure, where the Moore-Penrose inverse is not required anymore (see \cref{sec:grad}), and helps achieve significant savings in computational cost.  

%% ==================================================================
\subsection{K-L expansion based discretization of the hydraulic conductivity spatial random field} \label{subsec:SpatialField_distreize}

Working with seepage flow data, the target physical parameter of interest is the hydraulic conductivity random field $k\left(\mathbf{z},\omega\right):\mathcal{D}\times\Omega\rightarrow\mathbb{R}$. To enforce positivity constraints, this field is represented as a log-normal spatial random field which is considered to share the following relationship with the Gaussian Process ${u}\left(\mathbf{z},\omega\right)$:
\begin{equation}
{u}\left(\mathbf{z},\omega \right)=\log_{10}\left(k\left(\mathbf{z},\omega\right)-k_{\min}\right). \label{eq:log_k}
\end{equation}
Here $k_{\min}$ is the lower bound on the hydraulic conductivity
that guarantees $k\left(\mathbf{z},\omega\right)>k_{\min}\geq0$. 
In this paper, prior information is encoded through the stationary Gaussian Process ${u}\left(\mathbf{z},\omega\right) \sim \mathcal{GP}(\overline{{u}}, C)$, where $\overline{{u}}(\mathbf{z})$ is the mean of ${u}$, and $C\left(\mathbf{z},\mathbf{z}^{*}\right)$ is a smooth Gaussian autocovariance kernel:
\begin{equation}
C\left(\mathbf{z},\mathbf{z}^{*}\right)=v\exp\left(\frac{-\left(\mathbf{z}-\mathbf{z}^{*}\right)^{\top}\left(\mathbf{z}-\mathbf{z}^{*}\right)}{2l^{2}}\right),\label{eq:gauss_kernel}
\end{equation}
where $v$ and $l$ are the scale and the correlation length, respectively. 

An $M_1$-term truncated K-L expansion is used to represent the random field ${u}\left(\mathbf{z},\omega\right)$, thereby enabling a reduction in dimensionality of the number of parameters to be estimated. Assume that a bounding domain $\mathcal{D}'$ can be constructed that bounds all possible realizations of $\mathcal{D}\left({}^{2}\bm{\uptheta}\right)$, then for the same mean 
$\overline{{u}}(\mathbf{z})$ and autocovariance $C\left(\mathbf{z},\mathbf{z}^{*}\right)$ functions now defined on $\mathcal{D}'$, using the domain independence property, the truncated K-L expansion of ${u}\left(\mathbf{z},\omega\right)$ at all points $\mathbf{z} \in \mathcal{D}\left({}^{2}\bm{\uptheta}\right) \cap \mathcal{D}'$ can be approximated as
\begin{equation}
{u}\left(\mathbf{z},\omega\right) = {u}\left(\mathbf{z},{}^{1}\bm{\uptheta}\left( \omega \right)\right) \approx \overline{u}\left(\mathbf{z}\right)+\sum_{i=1}^{M_1}\sqrt{\lambda'_{i}}\phi'_{i}\left(\mathbf{z}\right){}^{1}\theta_{i}\left(\omega\right).\label{eq:log_KL}
\end{equation}
Here $\lambda'_{i}$ and $\phi'_{i}\left(\mathbf{z}\right)$
are computed with respect to the IEVP \cref{eq:IEVP} defined on the bounding domain $\mathcal{D}'$,
$^{1}\bm{\uptheta}\left( \omega \right)=\left[^{1}\theta_{1}\left(\omega\right),\ldots,{}^{1}\theta_{M_1}\left(\omega\right)\right]^{\top}$
is a standard normal random vector, and 
${u}\left(\mathbf{z},{}^{1}\bm{\uptheta}\left( \omega \right)\right)$ is a new notation for $u$, introduced to emphasize its parameterization by $^{1}\bm{\uptheta}$ \cite{marzouk2009}. For completeness, considering \cref{eq:log_k} and \cref{eq:log_KL}, the log-normal random field 
$k\left(\mathbf{z},\omega\right)$ can be restated as $k\left(\mathbf{z},{}^{1}\bm{\uptheta}\left( \omega \right)\right)$ and is given as
\begin{equation}
k\left(\mathbf{z},{}^{1}\bm{\uptheta}\right) =
k_{\min}+10^{{u}\left(\mathbf{z},{}^{1}\bm{\uptheta}\right)}.\label{eq:k(z)}
\end{equation}
The eigenvalues $\lambda'_{i}$ and eigenfunctions $\phi'_{i}\left(\mathbf{z}\right)$ of the IEVP are obtained numerically by the Nyström method \cite{betz2014,press2007numerical}. Using a Gaussian quadrature scheme for numerical integration, \cref{eq:IEVP} on $\mathcal{D}'$ can be approximated as
\begin{equation}
\sum_{j=1}^{N}w_{j}C\left(\mathbf{z},\mathbf{z}'_{j}\right)\phi'_{i}\left(\mathbf{z}'_{j}\right)=\lambda'_{i}\phi'_{i}\left(\mathbf{z}\right),\label{eq:2_Nys1}
\end{equation}
where $\left\{ \mathbf{z}'_{j}\right\} _{j=1}^{N}\subset\mathcal{D}'$
are integration points, and $\left\{ w_{j}\right\} _{j=1}^{N}\subset\mathbb{R}^{+}$ are corresponding integration weights. The method proceeds by requiring that \cref{eq:2_Nys1} is satisfied for $\mathbf{z}=\mathbf{z}'_{n}$ $\left(n=1,\ldots,N\right)$, i.e.,
\begin{equation}
\sum_{j=1}^{N}w_{j}C\left(\mathbf{z}'_{n},\mathbf{z}'_{j}\right)\phi'_{i}\left(\mathbf{z}'_{j}\right)=\lambda'_{i}\phi'_{i}\left(\mathbf{z}'_{n}\right),\quad\left(n=1,...,N\right).\label{eq:2_Nys2}
\end{equation}
The system of $N$ equations in \cref{eq:2_Nys2} can be written in a matrix form as
\begin{equation}
\mathbf{C}\mathbf{W}\mathbf{f}_{i}=\lambda'_{i}\mathbf{f}_{i}, \label{eq:2_eigen1}
\end{equation}
where $\mathbf{C}\in\mathbb{R}^{N\times N}$ is
a symmetric positive semi-definite matrix with elements $c_{nj}=C\left(\mathbf{z}'_{n},\mathbf{z}'_{j}\right)$,
$\mathbf{W}\in\mathbb{R}^{N\times N}$ is a diagonal matrix with elements
$W_{nj}=w_{j}\delta_{nj}$, and $\mathbf{f}_{i}\in\mathbb{R}^{N}$
is a vector whose $n$th entry is $f_{i,n}=\phi'_{i}\left(\mathbf{z}'_{n}\right)$.
Left multiplying \cref{eq:2_eigen1} by $\mathbf{W}^{\frac{1}{2}}$
with elements $W_{nj}^{\frac{1}{2}}=\sqrt{w_{j}\delta_{nj}}$, a matrix eigenvalue problem is obtained as follows:
\begin{equation}
\mathbf{B}\mathbf{g}_{i}=\lambda'_{i}\mathbf{g}_{i},\label{eq:2_eigen2}
\end{equation}
where $\mathbf{B}=\mathbf{W}^{\frac{1}{2}}\mathbf{C}\mathbf{W}^{\frac{1}{2}}$,
$\mathbf{g}_{i}=\mathbf{W}^{\frac{1}{2}}\mathbf{f}_{i}$.
Since $\mathbf{B}$ is a symmetric positive semi-definite matrix, $\lambda'_{i}\in\mathbb{R}_{0}^{+}$ and $\mathbf{g}_{i}\cdot\mathbf{g}_{j}=\delta_{ij}$
are satisfied. Noting that $\phi'_{i}\left(\mathbf{z}'_{j}\right)=\frac{1}{\sqrt{w_{j}}}g_{i,j}$ and using \cref{eq:2_Nys1}, the Nyström interpolation formula for the eigenfunction $\phi'_{i}\left(\mathbf{z}\right)$ can be obtained as
\begin{equation}
\phi'_{i}\left(\mathbf{z}\right)=\frac{1}{\lambda'_{i}}\sum_{j=1}^{N}\sqrt{w_{j}}g_{i,j}C\left(\mathbf{z},\mathbf{z}'_{j}\right),\label{eq:2_eigenfun}
\end{equation}
where $g_{i,j}$ is the $j$th entry of
$\mathbf{g}_{i}$.

\begin{figure}
\centering
\includegraphics{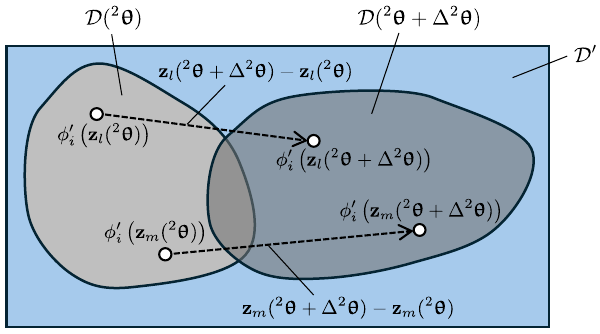}
\caption{\label{fig:eigenfunc}
Change in the location of evaluation of the eigenfunction $\phi'_{i}$ as the domain $\mathcal{D}\left( ^2\bm{\uptheta} \right)$ is updated due to an update in the geometry parameters $^2\bm{\uptheta} \rightarrow ^2\bm{\uptheta}+\Delta^2\bm{\uptheta}$. The shape of eigenfunctions $\phi'_{i}$ remains fixed on the bounding domain $\mathcal{D}'$.}
\end{figure}

In the discussion above, it must be emphasized that ${u}\left(\mathbf{z}_l\left( ^2\bm{\uptheta} \right),{}^{1}\bm{\uptheta}\right)$, the value of the random field at the point with coordinates $\mathbf{z}_l\left( ^2\bm{\uptheta} \right)$, depends on the geometry parameters ${}^{2}\bm{\uptheta}$ in the simultaneous estimation method.
This is because the point coordinates $\mathbf{z}_l\left( ^2\bm{\uptheta} \right)$, at which the eigenfunctions $\phi'_{i} \left( \cdot \right)$ are evaluated, move as the computational domain $\mathcal{D}\left({}^{2}\bm{\uptheta}\right)$ is updated (see \cref{fig:eigenfunc}). In this paper, the mesh updates corresponding to the computational domain, are done from a fixed reference domain to maintain reversibility of the HMC algorithm \cite{koch2020b}.
When $^2\bm{\uptheta}$ is updated to $^2\bm{\uptheta} + \Delta^2\bm{\uptheta}$, the value of $\phi'_{i}\left( \cdot \right)$ at $\mathbf{z}_l\left( ^2\bm{\uptheta} \right)$ (i.e., $\phi'_{i} \left( \mathbf{z}_l\left( ^2\bm{\uptheta} \right) \right)$) changes to $\phi'_{i} \left( \mathbf{z}_l\left(^2\bm{\uptheta} + \Delta^2\bm{\uptheta} \right) \right)$ because the evaluation point moves from $\mathbf{z}_l\left( ^2\bm{\uptheta} \right)$ to $\mathbf{z}_l\left(^2\bm{\uptheta} + \Delta^2\bm{\uptheta} \right)$.
In other words, even though 
%the eigenfunctions remain invariant in space,
the shape of eigenfunctions remains fixed,
the value $\phi'_{i} \left( \mathbf{z}_l\left( ^2\bm{\uptheta} \right) \right)$ depends on the geometry parameters $^{2}\bm{\uptheta}$. This dependence must be accounted for in the computation of gradients w.r.t the geometry parameters (see \cref{sec:grad}). It is worth mentioning that these gradients correspond to the advection term in the material derivative when considering $^2\bm{\uptheta} $ as time.

%% ==================================================================
%% =========================== Section 3 ============================
%% ==================================================================

%SECTION3
\section{Bayesian inference}\label{sec:Bayes}

%% ==================================================================
\subsection{Forward and observation models}\label{subsec:forward_model}
Consider a domain $\mathcal{D}' \subset \mathbb{R}^d$ that contains a soil domain $\mathcal{D}\left({}^2\bm{\uptheta}\right)$ (i.e., $\mathcal{D}\left({}^2\bm{\uptheta}\right) \subset \mathcal{D}'$).   
The steady seepage flow (ignoring transient effects) through $\mathcal{D}\left({}^2\bm{\uptheta}\right)$, governed by Darcy's law and the continuity equation, is written as
\begin{equation}
\mathbf{v}\left(\mathbf{z}, {}^1\bm{\uptheta} \right)=-k\left(\mathbf{z}, {}^1\bm{\uptheta}\right)\nabla h\left(\mathbf{z}, {}^1\bm{\uptheta} \right),\label{eq:govern1}
\end{equation}
\begin{equation}
\nabla\cdot\mathbf{v}\left(\mathbf{z}, {}^1\bm{\uptheta} \right) = Q\left(\mathbf{z}\right),\label{eq:govern2}
\end{equation}
where $\mathbf{z}\in\mathcal{D}\left({}^2\bm{\uptheta}\right)$,
%, which changes through updates of $\mathcal{D}\left({}^2\bm{\uptheta}\right)$,
$h\left(\mathbf{z}, {}^1\bm{\uptheta} \right)$ is the total hydraulic head, $\mathbf{v}\left(\mathbf{z}, {}^1\bm{\uptheta} \right)$ is the seepage flow velocity, $Q\left(\mathbf{z}\right)$ is the source term, and $k\left(\mathbf{z}, {}^1\bm{\uptheta}\right)$ is the hydraulic conductivity. The boundaries are classified as Dirichlet boundary $\Gamma_{\mathrm{D}}$ or Neumann boundary $\Gamma_\mathrm{N}$ (i.e., $\partial\mathcal{D}=\Gamma_{\mathrm{D}} \cup \Gamma_\mathrm{N}$, $\emptyset=\Gamma_\mathrm{D} \cap \Gamma_\mathrm{N}$), with the boundary conditions given as
\begin{equation}
h\left(\mathbf{z}\right)=h_{0}\left(\mathbf{z}\right)\quad\mathbf{z}\in\Gamma_{\mathrm D},
\end{equation}
\begin{equation}
\mathbf{v}\left(\mathbf{z} \right)\cdot\mathbf{n} =q_{n}\left(\mathbf{z}\right)\quad\mathbf{z}\in\Gamma_{\mathrm{N}}. 
\end{equation}
The term $h_{0}(\mathbf{z})$ is a known hydraulic head, $\mathbf{n} \in \mathbb{R}^d$ is the outward unit vector normal to $\Gamma_\mathrm{N}$ and $q_{n}(\mathbf{z})$ is the known flux on $\Gamma_\mathrm{N}$.

Consider a finite element discretization of the domain $\mathcal{D}\left({}^2\bm{\uptheta}\right)$ containing $n_{\mathrm{nd}}$ nodes. Taking into account that nodes are moved to conform to updated domains $\mathcal{D}\left({}^2\bm{\uptheta}\right)$ during inversion, the nodal coordinates matrix can be represented as $\mathbf{Z}\left({}^2\bm{\uptheta}\right) \in \mathbb{R}^{n_\mathrm{nd} \times d}$. Therefore, the discretized form of \cref{eq:govern1} and \cref{eq:govern2} at $\mathbf{Z}\left({}^2\bm{\uptheta}\right)$ depends not only on ${}^1\bm{\uptheta}$ but also on ${}^2\bm{\uptheta}$ and is given as
\begin{equation}
\mathbf{K}(\bm{\uptheta})\mathbf{h}(\bm{\uptheta})= \mathbf{q}(\bm{\uptheta}),\label{eq:fem}
\end{equation}
where $\mathbf{h},\mathbf{q}\in\mathbb{R}^{n_\mathrm{nd}}$ are the global discretized hydraulic head and nodal flux vectors, and $\mathbf{K}$ is the global hydraulic conductivity matrix. The elemental hydraulic conductivity matrix is 
\begin{equation}
\mathbf{K}_{e}(\bm{\uptheta})= \int_{\mathcal{\mathcal{D}}_{e}}
\mathbf{G}({}^2\bm{\uptheta})^{\top}k_{e}(\bm{\uptheta})\mathbf{G}({}^2\bm{\uptheta})\left|\mathbf{J}_{e}({}^2\bm{\uptheta})\right|\mathrm{d}\mathcal{D}_{e},\label{eq:elemat}
\end{equation}
where $k_{e}(\bm{\uptheta}) = k\left(\mathbf{z}_e^\mathrm{e}({}^2\bm{\uptheta}), {}^1\bm{\uptheta} \right)$
is the hydraulic conductivity in the element $e$ with the central coordinates $\mathbf{z}_e^\mathrm{e}({}^2\bm{\uptheta})$, $\mathbf{G}$ is a matrix containing derivatives of the elemental shape functions, $\left|\mathbf{J}_{e}\right|$ is the determinant of the elemental Jacobian matrix, and $\mathcal{D}_{e}$ is the region occupied by the element $e$ in isoparametric space. Note that the dependence of $\mathbf{G}$ and $\left|\mathbf{J}_{e}\right|$ on ${}^2\bm{\uptheta}$ arises from their construction using the nodal coordinates matrix for the element $e$, which is a submatrix of $\mathbf{Z}\left({}^2\bm{\uptheta}\right)$.

Observations used in the inverse analysis consist of measurements of hydraulic head and flux values measured at discrete locations in $\mathcal{D}$. Let $\mathbf{x}_{t}(\bm{\uptheta})=\left[\mathbf{h}_{t}(\bm{\uptheta}),\mathbf{q}_{t}(\bm{\uptheta})\right]^{\top}$ be the state vector, and $\mathbf{y}_{t}$ be observations masked by Gaussian noise $\mathbf{r}_{t}\sim\mathcal{N}\left(\mathbf{0},\mathbf{R}_{t}\right)$, where $\mathbf{R}_{t}$ is the covariance matrix. Then, the linear measurement model is given as
\begin{equation}
\mathbf{y}_{t}=\mathbf{H}\mathbf{x}_{t}(\bm{\uptheta})+\mathbf{r}_{t},\label{eq:error}
\end{equation}
where $\mathbf{H}$ is the measurement model matrix. In \cref{eq:error} the index $t=\{1,\ldots,n\}$ refers to independent observation data collected under $n$ different boundary conditions.

%% ==================================================================
\subsection{Bayes' theorem}
The parameter $\bm{\uptheta}$ to be estimated is approximated from the posterior probability distribution $p\left(\bm{\uptheta}|\mathbf{y}\right)$ given as 
\begin{equation}
p\left(\bm{\uptheta}|\mathbf{y}\right)=\frac{p\left(\mathbf{y}|\bm{\uptheta}\right)p\left(\bm{\uptheta}\right)}{\int p\left(\mathbf{y}|\bm{\uptheta}\right)p\left(\bm{\uptheta}\right)\mathrm{d}\bm{\uptheta}},\label{eq:bayes}
\end{equation}
where 
$\mathbf{y}$ refers to $\{ \mathbf{y}_{1}, \ldots, \mathbf{y}_{n} \}$,
$p(\mathbf{y}|\bm{\uptheta})$ is the likelihood, and $p\left(\bm{\uptheta}\right)$ is the prior distribution. Considering \cref{eq:error}, $p(\mathbf{y}|\bm{\uptheta})$ can be formulated as
\begin{equation}
p\left(\mathbf{y}|\bm{\uptheta}\right)=\prod_{t=1}^{n}p\left(\mathbf{y}_{t}|\bm{\uptheta}\right)=\prod_{t=1}^{n}\mathcal{N}\left(\mathbf{H}\mathbf{x}_{t}(\bm{\uptheta}),\mathbf{R}_{t}\right).\label{eq:likelihood}
\end{equation}
Noting that ${}^1\bm{\uptheta}$ is Gaussian (see \cref{eq:KL}), if the parameter ${}^2\bm{\uptheta}$ is also chosen to be Gaussian, then the prior distribution over $\bm{\uptheta}$ is given by the Gaussian density
\begin{equation}
p\left(\bm{\uptheta}\right)=\mathcal{N}\left(\bm{\upmu},\bm{\Sigma}_{\bm{\uptheta}}\right),\label{eq:prior}
\end{equation}
where $\bm{\upmu} \in \mathbb{R}^{M}$ is the mean, and $\bm{\Sigma}_{\bm{\uptheta}} \in \mathbb{R}^{M \times M}$ is the covariance matrix. These quantities are chosen apriori and are set according to the inversion method; in the spatial-field-only estimation method $M = M_1$, while in the simultaneous estimation method $M = M_1 + M_2$. Substituting \cref{eq:likelihood} and \cref{eq:prior} into \cref{eq:bayes}, the posterior density can be shown to be: 
\begin{equation}
p\left(\bm{\uptheta}|\mathbf{y}\right)\propto\exp\left(-\varphi\left(\bm{\uptheta}\right)\right),\label{eq:posterior}
\end{equation}
where 
\begin{equation}
\varphi\left(\bm{\uptheta}\right)
\equiv
\sum_{t=1}^{n}\frac{1}{2}\left(\mathbf{y}_{t}-\mathbf{Hx}_{t}\left(\bm{\uptheta}\right)\right)^{\top}\mathbf{R}_{t}^{-1}\left(\mathbf{y}_{t}-\mathbf{Hx}_{t}\left(\bm{\uptheta}\right)\right)+\frac{1}{2}\left(\bm{\uptheta}-\bm{\upmu}\right)^{\top}\bm{\Sigma_{\bm{\uptheta}}}^{-1}\left(\bm{\uptheta}-\bm{\upmu}\right).\label{eq:phi}
\end{equation}
Note that constant terms of $p\left(\bm{\uptheta}|\mathbf{y}\right)$ cancel out during the calculation of the acceptance probability in any MCMC algorithms, and it is sufficient to only consider $\bm{\uptheta}$-dependent terms of $p\left(\bm{\uptheta}|\mathbf{y}\right)$. 

%% ==================================================================
\subsection{Hamiltonian Monte Carlo}
The Hamiltonian Monte Carlo (HMC) algorithm is chosen to sample the posterior distribution of \cref{eq:posterior}. In HMC, auxiliary momentum variables $\mathbf{p} \in \mathbb{R}^{M}$ are introduced, and a joint distribution of $\bm{\uptheta}$ and $\mathbf{p}$ is defined as
\begin{equation}
p\left(\bm{\uptheta},\mathbf{p}\right)=p\left(\mathbf{p}|\bm{\uptheta}\right)p\left(\bm{\uptheta}|\mathbf{y}\right).\label{eq:jointdist}
\end{equation}
We make a simple choice for the probability distribution of $\mathbf{p}$ which is independent of $\bm{\uptheta}$, i.e., $\mathbf{p} \sim \mathcal{N}\left(\mathbf{0},\bm{\mathcal{M}}\right)$. Thus, $p\left(\mathbf{p}|\bm{\uptheta}\right)$ can be expressed as
\begin{equation}
p\left(\mathbf{p}|\bm{\uptheta}\right)=p\left(\mathbf{p}\right)\propto\exp\left(-\mathcal{K}\left(\mathbf{p}\right)\right),\label{eq:p_dist}
\end{equation}
where
\begin{equation}
\mathcal{K}\left(\mathbf{p}\right)=\frac{1}{2}\mathbf{p}^{\top}\bm{\mathcal{M}}^{-1}\mathbf{p}.
\end{equation}
Here, $\bm{\mathcal{M}}$ is a symmetric positive-definite mass matrix. Considering \cref{eq:posterior,eq:jointdist,eq:p_dist}, the joint distribution $p\left(\bm{\uptheta},\mathbf{p}\right)$ can be represented as
\begin{equation}
p\left(\bm{\uptheta},\mathbf{p}\right)\propto\exp\left(-\mathcal{\mathcal{H}}\left(\bm{\uptheta},\mathbf{p}\right)\right),
\end{equation}
where $\mathcal{\mathcal{H}}\left(\bm{\uptheta},\mathbf{p}\right)$
is the Hamiltonian and defined as
\begin{equation}
\mathcal{\mathcal{H}}\left(\bm{\uptheta},\mathbf{p}\right)=\mathcal{K}\left(\mathbf{p}\right)+\varphi\left(\bm{\uptheta}\right).
\end{equation}
In this context, $\mathcal{K}\left(\mathbf{p}\right)$ is called the kinetic energy, and $\varphi\left(\bm{\uptheta}\right)$ is called the potential energy. The definition of the Hamiltonian enables the generation of deterministic trajectories of $\bm{\uptheta}$ through Hamiltonian dynamics. 
These equations are often solved numerically by the leapfrog method, which satisfies time reversibility and volume conservation properties. This method consists of the following three steps:
\begin{equation}
\mathbf{p}\left(t+\frac{\varepsilon}{2}\right) =\mathbf{p}\left(t\right)-\frac{\varepsilon}{2}\frac{\partial\varphi\left(\bm{\uptheta}\left(t\right)\right)}{\partial\bm{\uptheta}} ,
\end{equation}
\begin{equation}
\bm{\uptheta}\left(t+\varepsilon\right)  =\bm{\uptheta}\left(t\right)+\varepsilon\bm{\mathcal{M}}^{-1}\mathbf{p}\left(t+\frac{\varepsilon}{2}\right),
\end{equation}
\begin{equation}
\mathbf{p}\left(t+\varepsilon\right) =\mathbf{p}\left(t+\frac{\varepsilon}{2}\right)-\frac{\varepsilon}{2}\frac{\partial\varphi\left(\bm{\uptheta}\left(t+\varepsilon\right)\right)}{\partial\bm{\uptheta}}. 
\end{equation}
Starting with a random draw $\mathbf{p}^j \sim \mathcal{N}\left(\mathbf{0},\bm{\mathcal{M}}\right)$, the current $j$th sample $\left(\bm{\uptheta}^{j},\mathbf{p}^{j}\right)$ is updated to the $\left(j+1\right)$th sample candidate $\left(\bm{\uptheta}',\mathbf{p}'\right)$ by applying the leapfrog steps $L$ times, with step-size $\varepsilon$. Whether $\left(\bm{\uptheta}',\mathbf{p}'\right)$
is accepted or rejected as the $\left(j+1\right)$th sample is determined by the Metropolis-Hastings acceptance criteria \cite{afshar2021}:
\begin{eqnarray}
\alpha_{j+1}\left(\left(\bm{\uptheta}^{j},\mathbf{p}^{j}\right),\left(\bm{\uptheta}',\mathbf{p}'\right)\right) =\min\left\{ 1,\frac{p\left(\bm{\uptheta}',\mathbf{p}'\right)}{p\left(\bm{\uptheta}^{j},\mathbf{p}^{j}\right)}\left|\frac{\partial\left(\bm{\uptheta}',\mathbf{p}'\right)}{\partial\left(\bm{\uptheta}^{j},\mathbf{p}^{j}\right)}\right|\right\} \nonumber \\
 =\min\left\{ 1,\exp\left(-\mathcal{\mathcal{H}}\left(\bm{\uptheta}',\mathbf{p}'\right)+\mathcal{H}\left(\bm{\uptheta}^{j},\mathbf{p}^{j}\right)\right)\right\},
\end{eqnarray}
where $\left|\frac{\partial\left(\bm{\uptheta}',\mathbf{p}'\right)}{\partial\left(\bm{\uptheta}^{j},\mathbf{p}^{j}\right)}\right|$
is the Jacobian for the conversion from $\left(\bm{\uptheta}^{j},\mathbf{p}^{j}\right)$
to $\left(\bm{\uptheta}',\mathbf{p}'\right)$, and $\left|\frac{\partial\left(\bm{\uptheta}',\mathbf{p}'\right)}{\partial\left(\bm{\uptheta}^{j},\mathbf{p}^{j}\right)}\right|=1$
in the leapfrog due to volume conservation. The choice of the
parameters $\varepsilon$, $L$, and $\bm{\mathcal{M}}$ in the leapfrog scheme affects the sampling efficiency of HMC. Hence, methods that automatically tune these parameters are useful. In this study, we use an adaptation procedure similar to Stan \cite{carpenter2017}. In detail, the Dual Averaging scheme \cite{nesterov2009} is used for tuning $\varepsilon$, the No-U-Turn Sampler \cite{hoffman2014} is used for tuning $L$, and the adaptation of $\bm{\mathcal{M}}$ is done according to section 4.2.1 in \cite{betancourt2018}.

%% ==================================================================
%% =========================== Section 4 ============================
%% ==================================================================

\section{Gradient computation} \label{sec:grad}

The gradient of the negative log of the posterior density w.r.t the parameter $\bm{\uptheta}$, required in the leapfrog steps, is obtained through differentiation of \cref{eq:phi} and is given as
\begin{equation}
\frac{\partial\varphi(\bm{\uptheta})}{\partial\bm{\uptheta}}=\sum_{t=1}^{n}-\left(\mathbf{y}_{t}-\mathbf{Hx}_{t}\left(\bm{\uptheta}\right)\right)^{\top}\mathbf{R}_{t}^{-1}\left(\mathbf{H}\frac{\partial\mathbf{x}_{t}(\bm{\uptheta})}{\partial\bm{\uptheta}}\right)+(\bm{\uptheta}-\bm{\upmu})^{\top}\bm{\Sigma_{\bm{\uptheta}}}^{-1}, \label{eq:ddm}
\end{equation}
where $\frac{\partial\mathbf{x}_{t}\left(\bm{\uptheta}\right)}{\partial\bm{\uptheta}}=\left[\frac{\partial\mathbf{h}_{t}\left(\bm{\uptheta}\right)}{\partial\bm{\uptheta}},\frac{\partial\mathbf{q}_{t}\left(\bm{\uptheta}\right)}{\partial\bm{\uptheta}}\right]^{\top}$. \cref{eq:ddm} refers to the direct differentiation method (DDM) and involves the expensive computation of the possibly large state vectors w.r.t $\bm{\uptheta}$. This cost can be reduced through the adjoint method (AM), details of which can be found in \cite{koch2020a, plessix2006review}. While the direct computation of $\frac{\partial\mathbf{x}_{t}\left(\bm{\uptheta}\right)}{\partial\bm{\uptheta}}$ can be avoided if the AM method is used, the term $\frac{\partial\mathbf{K}}{\partial\bm{\uptheta}}$ has to be computed both in DDM and AM. These derivatives of the FEM global hydraulic conductivity matrix $\frac{\partial\mathbf{K}}{\partial\bm{\uptheta}}=\sum_{e}\frac{\partial\mathbf{K}_{e}}{\partial\bm{\uptheta}}$, where $\frac{\partial\mathbf{K}_{e}}{\partial\bm{\uptheta}}$ is the gradient of the elemental hydraulic conductivity matrix, can be obtained from \cref{eq:elemat}. 
%\begin{equation}
%\frac{\partial\mathbf{K}}{\partial\bm{\uptheta}}\mathbf{h}_{t}+\mathbf{K}\frac{\partial\mathbf{h}_{t}}{\partial\bm{\uptheta}}=\frac{\partial \mathbf{q}_{t}} {\partial\bm{\uptheta}},
%\end{equation}
This gradient consists of two derivatives w.r.t the spatial field and the geometry parameters (i.e., $\frac{\partial\mathbf{K}_{e}}{\partial{}^{1}\bm{\uptheta}}$ and $\frac{\partial\mathbf{K}_{e}}{\partial{}^{2}\bm{\uptheta}}$, respectively), which are given as
\begin{equation}
\frac{\partial\mathbf{K}_{e}}{\partial{}^{1}\bm{\uptheta}}=\int_{\mathcal{\mathcal{D}}_{e}}\mathbf{G}^{\top}\frac{\partial k_{e}}{\partial{}^{1}\bm{\uptheta}}\mathbf{G}\left|\mathbf{J}_{e}\right|\mathrm{d}\mathcal{D}_{e}, \label{eq:dK_dtheta_1}
\end{equation}

\begin{equation}
\begin{split}
\frac{\partial\mathbf{K}_{e}}{\partial{}^{2}\bm{\uptheta}}=
\int_{\mathcal{\mathcal{D}}_{e}}&\left( \frac{\partial\mathbf{G}^{\top}}{\partial{}^{2}\bm{\uptheta}}k_{e}\mathbf{G}\left|\mathbf{J}_{e}\right|
+ \mathbf{G}^{\top}k_{e}\frac{\partial\mathbf{G}}{\partial{}^{2}\bm{\uptheta}}\left|\mathbf{J}_{e}\right| \right. \\
&+ \left. \mathbf{G}^{\top}k_{e}\mathbf{G}\frac{\partial\left|\mathbf{J}_{e}\right|}{\partial{}^{2}\bm{\uptheta}}
+\mathbf{G}^{\top}\frac{\partial k_{e}}{\partial{}^{2}\bm{\uptheta}}\mathbf{G}\left|\mathbf{J}_{e}\right|\right)\mathrm{d}\mathcal{D}_{e}. 
\end{split}
\label{eq:dK_dtheta_2}
\end{equation}
The elemental domains in isoparametric space $\mathcal{D}_{e}$ map to the domains $\mathcal{D}({}^{2}\bm{\uptheta})$, which are updated for every momentum update in the leapfrog scheme. The derivatives $\frac{\partial\mathbf{G}}{\partial{}^{2}\bm{\uptheta}}$ and $\frac{\partial\mathbf{\left|\mathbf{J}_{e}\right|}}{\partial{}^{2}\bm{\uptheta}}$ are common in shape optimization literature and calculated in the same way as \cite{koch2021}. For completeness, these are given as
\begin{equation}
\frac{\partial\mathbf{G}}{\partial{}^{2}\bm{\uptheta}}=-\mathbf{G}\frac{\partial\mathbf{Z}_{e}}{\partial{}^{2}\bm{\uptheta}}\mathbf{G},
\end{equation}

\begin{equation}
\frac{\partial\mathbf{\left|\mathbf{J}_{e}\right|}}{\partial{}^{2}\bm{\uptheta}}=\mathbf{\left|\mathbf{J}_{e}\right|}\mathrm{ tr}\left(\mathbf{G}\frac{\partial\mathbf{Z}_{e}}{\partial{}^{2}\bm{\uptheta}}\right),
\end{equation}
where $\mathbf{Z}_{e}$ is the nodal coordinate matrix for element
$e$. 

The gradient of the hydraulic conductivity random field (w.r.t spatial field parameters $\frac{\partial k}{\partial{}^{1}\bm{\uptheta}}$ and geometry parameters $\frac{\partial k}{\partial{}^{2}\bm{\uptheta}}$)
appears in \cref{eq:dK_dtheta_1} and \cref{eq:dK_dtheta_2}, respectively. The combined vector $\frac{\partial k}{\partial\bm{\uptheta}}$ can be obtained through a direct differentiation of \cref{eq:k(z)} as
\begin{equation}
\frac{\partial k\left(\mathbf{z}\left({}^{2}\bm{\uptheta}\right),{}^{1}\bm{\uptheta}\right)}{\partial\bm{\uptheta}}=\frac{\partial{u}\left(\mathbf{z}\left({}^{2}\bm{\uptheta}\right),{}^{1}\bm{\uptheta}\right)}{\partial\bm{\uptheta}}10^{{u}\left(\mathbf{z}\left({}^{2}\bm{\uptheta}\right),{}^{1}\bm{\uptheta}\right)}\ln10.
\end{equation}
From the discussion in \cref{subsec:SpatialField_distreize}, it is clear that 
%the Gaussian random field 
${u}\left(\mathbf{z}\left({}^{2}\bm{\uptheta}\right),{}^{1}\bm{\uptheta}\right)$ depends on both ${}^{1}\bm{\uptheta}$ and ${}^{2}\bm{\uptheta}$. The derivative $\frac{\partial{u}}{\partial\bm{\uptheta}}$ can be obtained from a simple differentiation of \cref{eq:log_KL}, and its components $\frac{\partial{u}}{\partial{}^{1}\bm{\uptheta}}$
and $\frac{\partial{u}}{\partial{}^{2}\bm{\uptheta}}$ are
\begin{equation}
\frac{\partial{u}\left(\mathbf{z}\left({}^{2}\bm{\uptheta}\right),{}^{1}\bm{\uptheta}\right)}{\partial{}^{1}\theta_{i}}=\sqrt{\lambda'_{i}}\phi'_{i}\left(\mathbf{z}\left({}^{2}\bm{\uptheta}\right)\right),
\end{equation}

\begin{equation}
\frac{\partial{u}\left(\mathbf{z}\left({}^{2}\bm{\uptheta}\right),{}^{1}\bm{\uptheta}\right)}{\partial{}^{2}\bm{\uptheta}}=\sum_{i=1}^{M_1}\sqrt{\lambda'_{i}}{}^{1}\theta_{i}\frac{\partial\phi'_{i}\left(\mathbf{z}\left({}^{2}\bm{\uptheta}\right)\right)}{\partial{}^{2}\bm{\uptheta}}.
\end{equation}
For a numerical computation of the IEVP in \cref{eq:IEVP} using the Nyström method, the shape derivatives of the eigenfunctions $\frac{\partial\phi'_{i}}{\partial{}^{2}\bm{\uptheta}}$ are obtained from \cref{eq:2_eigenfun} as
\begin{equation}
\frac{\partial\phi'_{i}\left(\mathbf{z}\left({}^{2}\bm{\uptheta}\right)\right)}{\partial{}^{2}\bm{\uptheta}}=\frac{1}{\lambda'_{i}}\sum_{j=1}^{N}\sqrt{w_{j}}g_{i,j}\frac{\partial C\left(\mathbf{z}\left({}^{2}\bm{\uptheta}\right),\mathbf{z}'_{j}\right)}{\partial{}^{2}\bm{\uptheta}}. \label{eq:dPhi_d2theta}
\end{equation}
Here $\frac{\partial C\left(\mathbf{z}\left({}^{2}\bm{\uptheta}\right),\mathbf{z}'_{j}\right)}{\partial{}^{2}\bm{\uptheta}}$
is the derivative of the autocovariance function in \cref{eq:gauss_kernel}. Seeing that the
integration points $\left\{ \mathbf{z}'_{j}\right\} _{j=1}^{N}$ employed in the Nyström method (see \cref{subsec:SpatialField_distreize}) are defined on
$\mathcal{D}'$ and are independent of $^{2}\bm{\uptheta}$, $\frac{\partial C\left(\mathbf{z},\mathbf{z}_{j}\right)}{\partial{}^{2}\bm{\uptheta}}$ is given as
\begin{equation}
\frac{\partial C\left(\mathbf{z}\left({}^{2}\bm{\uptheta}\right),\mathbf{z}'_{j}\right)}{\partial{}^{2}\bm{\uptheta}}= 
\frac{\partial C\left(\mathbf{z}\left({}^{2}\bm{\uptheta}\right),\mathbf{z}'_{j}\right)}{\partial{}\mathbf{z}}\frac{\partial\mathbf{z}\left({}^{2}\bm{\uptheta}\right)}{\partial{}^{2}\bm{\uptheta}},
\end{equation}
where 
\begin{equation}
\frac{\partial C\left(\mathbf{z},\mathbf{z}'_{j}\right)}{\partial{}\mathbf{z}}=
-\frac{v}{l^{2}}\left(\mathbf{z}-\mathbf{z}'_{j}\right)^{\top}\exp\left(\frac{-\left(\mathbf{z}-\mathbf{z}'_{j}\right)^{\top}\left(\mathbf{z}-\mathbf{z}'_{j}\right)}{2l^{2}}\right).
\end{equation}

When the domain independence property of the K-L expansion is not used in the simultaneous estimation method of Koch et al. (2021) \cite{koch2021}, the derivatives of the K-L expansion require the gradient calculations associated with the IEVP in \cref{eq:IEVP}. Specifically, the shape derivatives of the matrix eigenvalue problem in \cref{eq:2_eigen2}, which is the discrete form of the IEVP, must be computed. This process involves the computation of the gradients of the eigenvalues and eigenfunctions w.r.t the geometry parameters, the latter of which involves the expensive Moore-Penrose inverse. The same problem arises in random field discretization through the discrete K-L expansion \cite{koch2021} and is discussed in \ref{app:discreteKL}. Furthermore, a new IEVP would be set up for every update of the domain $\mathcal{D}({}^{2}\bm{\uptheta})$, necessitating the computation of these expensive gradients in every step of HMC. Implementation of the domain independence property of the K-L expansion obviates the need to compute the shape derivatives of the eigenvalues. Although the shape derivatives of the eigenfunctions still have to be computed, these derivatives (see \cref{eq:dPhi_d2theta}) only involve terms already calculated in the Nyström method and the cheap computation of the derivative of the autocovariance function.

%% ==================================================================
%% =========================== Section 5 ============================
%% ==================================================================
\section{Numerical experiments}\label{sec:NumExp}

The inversion is carried out using three methods: the spatial-field-only estimation method \cite{koch2020a}, the simultaneous estimation method of Koch et al. (2021) \cite{koch2021}, and the proposed simultaneous estimation method using the domain-independence property of the K-L expansion. These methods are applied to two seepage flow problems containing geometric features of interest. In the simultaneous method of Koch et al. (2021) (see \ref{app:discreteKL}), the discretization of the random field is done through the discrete K-L expansion \cite{li2006efficient}. The first problem is a 1D vertical seepage flow problem through a 3-layered soil where the hydraulic conductivity in the three layers as well as the actual location of the layer interfaces is unknown. The second problem involves the determination of the hydraulic conductivity in a domain where a pipe of unknown size and location is known to be present. Inversion is performed using HMC, and the program for all three methods is written in Julia based on the AdvancedHMC.jl \cite{xu2020advancedhmc} package. The impact of the prior correlation length on inversion results is examined in the following sections. Furthermore, comments are made on the computation savings achieved through the implementation of the domain independence property.

%% ==================================================================
\subsection{1D seepage flow problem}
\begin{figure}[t]
\centering
\includegraphics[width=1.0\textwidth]{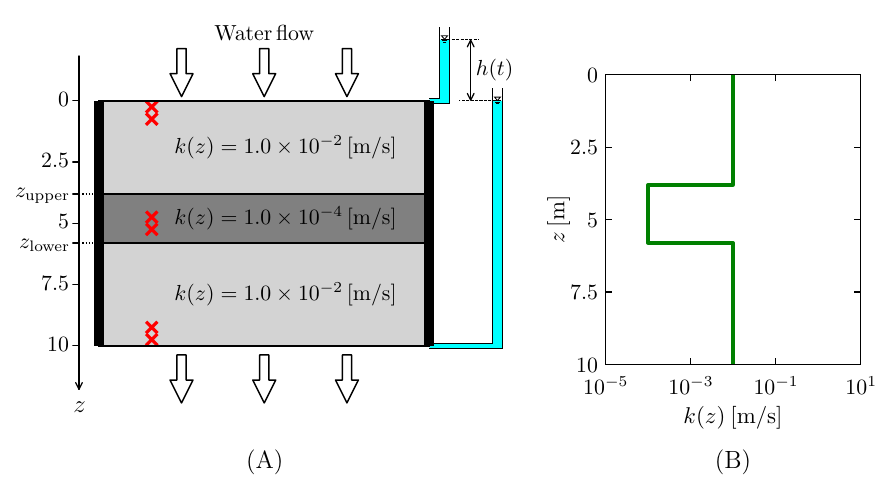}
\caption{\label{fig:Setup}(A) 
Target domain with a thin clay layer sandwiched in between two homogeneous sandy layers. Observations consist of hydraulic heads 
at 6 red cross mark points ($z = \{0.25, 0.75, 4.75, 5.25, 9.25, 9.75\}$) and inflow rate at the top. The boundary coordinates of the clay layer are represented by $z_{\mathrm{upper}}$ and $z_{\mathrm{lower}}$. 
(B) True hydraulic conductivity field.}
\end{figure}
A schematic of the 1D steady vertical seepage flow through three horizontal soil layers is shown in \cref{fig:Setup}(A). The target for estimation is the hydraulic conductivity spatial field in a domain $\mathcal{D}'(=\left\{z: 0 \leq z \leq 10 \right\})$, consisting of a thin clay seam in $\mathcal{D}_2(=\left\{z:3.8\leq z \leq 5.8 \right\})$ with hydraulic conductivity two orders of magnitude lower than that in the homogeneous sandy layers in the domain $\mathcal{D}_1 (= \mathcal{D}' \backslash \mathcal{D}_2)$ surrounding it. The true hydraulic conductivity field is shown in \cref{fig:Setup}(B). In addition to the hydraulic conductivity field, the locations of the top and bottom boundaries of the clay seam are unknown. Dirichlet boundary conditions (BCs) are implemented such that the hydraulic head at the ground surface varies as $ h(z=0,t)=0.1+0.01(t-1)$ for $t=\{1,...,31\}$, while $h(z=10,t) = 0$ is kept fixed. Observation data consists of the hydraulic head measured at 6 points (red cross marks in \cref{fig:Setup}(A)) and the water inflow rate measured at the top for 31 sets of BCs. These observations were obtained by adding Gaussian noise (mean is 0, standard deviation is 10\% of the true value) to the true values. The true values are obtained numerically by solving the forward problem for the true condition on a mesh with 200 equally spaced finite elements in $\mathcal{D}'$.

%% ==================================================================
\subsubsection{Comparison of estimated results} \label{subsubsec:1Dresult}
\begin{figure}[t]
\centering
\includegraphics[height=0.55\textwidth]{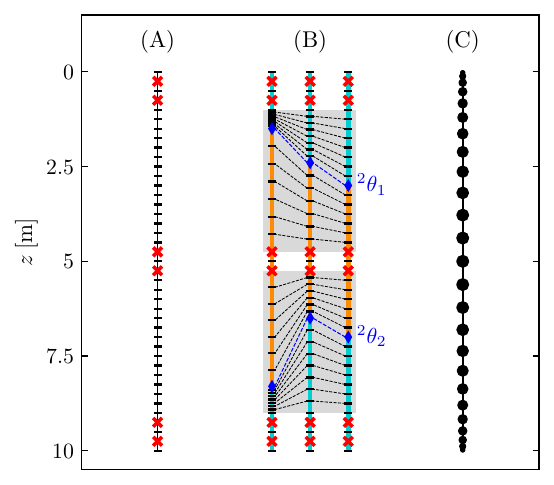}
\caption{\label{fig:mesh1D} (A) Mesh for the
spatial-field-only estimation. (B) Mesh for the simultaneous estimation. Only nodes in the gray area move, and observation points (red cross marks) are fixed. The cyan and orange lines correspond to $\mathcal{D}_1$ and $\mathcal{D}_2$, respectively. (C) Gauss quadrature points on the bounding domain $\mathcal{D}'$, where node size represents the magnitude of the Gauss weights.}
\end{figure}
%In all three methods of inversion, the log-normal hydraulic conductivity random field is discretized as $\mathbf{k}=\left[k\left(z^\mathrm{e}_{1}\right),\ldots,k\left(z^\mathrm{e}_{40}\right)\right]^{\top}$. These entries correspond to the hydraulic conductivity in 40 finite elements on $ 0 \leq z \leq 10$ (see \cref{fig:mesh1D}(A)).
In all three methods of inversion, the log-normal hydraulic conductivity random field is spatially discretized as $\mathbf{k}\left( \bm{\uptheta} \right) =\left[k_1\left( \bm{\uptheta} \right),\ldots,k_{40}\left( \bm{\uptheta} \right)\right]^{\top}$ for 40 finite elements on $ 0 \leq z \leq 10$ (see \cref{fig:mesh1D}(A)). Here, $k_e$ is the random hydraulic conductivity in the element $e$ with the central coordinate $z_e^{\mathrm{e}}$. In the spatial-field-only estimation method, $\mathbf{k}$ is represented through the K-L expansion of the Gaussian random field $u$ with the mean $\overline{u}=-3$ according to \cref{eq:log_k}. The K-L expansion is truncated at $M_1$ terms such that $\lambda_{M_1}>10^{-3}$ (see \cref{eq:truncated_KL}); thus, $\mathbf{k}$ is represented by the lower-dimensional vector $^{1}\bm{\uptheta}\in\mathbb{R}^{M_1}$, leading to the Gaussian prior $^{1}\bm{\uptheta} \sim \mathcal{N}\left(\mathbf{0},\mathbf{I}_{M_1}\right)$. The lower bound on $k$ is set to $k_{\mathrm{min}}=10^{-7}\, \mathrm{m/s}$. The prior scale $v$ of the Gaussian kernel is set as $v=1$, and three sets of prior correlation lengths $l=0.5,1.0,2.5\,\mathrm{m}$ are chosen for inversion.

In the simultaneous estimation methods, the hydraulic conductivity field is estimated separately on $\mathcal{D}_1$ and $\mathcal{D}_2$ (see \cref{fig:mesh1D}(B)), defined as
\begin{equation}
\begin{gathered}
\mathcal{D}_1({}^{2}\bm{\uptheta}) = 
\left\{ 0 < z < {}^{2}\theta_{1}, {}^{2}\theta_{2}< z < 10 \right\},\\
\mathcal{D}_2({}^{2}\bm{\uptheta}) = 
\left\{ {}^{2}\theta_{1} \leq z \leq {}^{2}\theta_{2} \right\},
\end{gathered}
\end{equation}
where the domain boundaries are parameterized through $^{2}\bm{\uptheta}=\left[^{2}\theta_1, ^{2}\theta_2 \right]^{\top}
=\left[z_{\mathrm{upper}}, z_{\mathrm{lower}}\right]^{\top}$. 
The mesh consists of 41 nodes (see \cref{fig:mesh1D}(B)), the position of each defined through the geometry parameters as 
\begin{equation}
z_{i}({}^{2}\bm{\uptheta})=\begin{cases}
0.25(i-1) & (1\leq i\leq5,20\leq i\leq22,37\leq i\leq41)\\
1+\frac{^{2}\theta_{1}-1}{8}(i-5) & (6\leq i\leq12)\\
^{2}\theta_{1} & (i=13)\\
^{2}\theta_{1}+\frac{4.75-^{2}\theta_{1}}{7}(i-13) & (14\leq i\leq19)\\
5.25+\frac{^{2}\theta_{2}-5.25}{7}(i-22) & (23\leq i\leq28)\\
^{2}\theta_{2} & (i=29)\\
^{2}\theta_{2}+\frac{9-^{2}\theta_{2}}{8}(i-29) & (30\leq i\leq36).
\end{cases} 
\label{eq:1Dprob_ndcoord}
\end{equation}
The map presented in \cref{eq:1Dprob_ndcoord} identifies a unique mesh configuration for each realization of ${}^{2}\bm{\uptheta}$ (see \cref{fig:mesh1D}(B)). This unique mapping guarantees reversibility as the finite element global hydraulic conductivity matrix $\mathbf{K}$ is constant for a particular realization of ${}^{2}\bm{\uptheta}$. Hence, updates from a reference configuration \cite{koch2020b} are not required in this simple example, and updates can be made from the previous HMC step. 
Non-physical mesh realizations of the mesh are restricted through the constraints $1.05<{}^{2}\theta_{1}<4.7$ and $5.3<{}^{2}\theta_{2}<8.95$.
Additionally, the observation nodes remain fixed in space.

For the separate domains $\mathcal{D}_1({}^{2}\bm{\uptheta})$ and $\mathcal{D}_2({}^{2}\bm{\uptheta})$, the hydraulic conductivity random field is divided into two parts:
\begin{equation}
k\left( z, {}^1\bm{\uptheta} \right)=
\begin{cases}
k_{\mathcal{D}_1}\left( z, {}^1\bm{\uptheta}_1 \right) & z\in\mathcal{D}_1({}^{2}\bm{\uptheta})\\
k_{\mathcal{D}_2}\left( z, {}^1\bm{\uptheta}_2 \right) & z\in\mathcal{D}_2({}^{2}\bm{\uptheta}),
\end{cases}\label{eq:2KL_for_1Dprob}
\end{equation}
Correspondingly, the hydraulic conductivity random vector $\mathbf{k} \left( \bm{\uptheta} \right)$ is also divided into two vectors:
\begin{equation}
\begin{gathered}
\begin{aligned}
\mathbf{k}_{\mathcal{D}_1}\left( \bm{\uptheta} \right) =&
\left[
k_{\mathcal{D}_1}\left(z^\mathrm{e}_{1}({}^{2}\bm{\uptheta}), {}^{1}\bm{\uptheta}_1\right),\ldots,k_{\mathcal{D}_1}\left(z^\mathrm{e}_{12}
({}^{2}\bm{\uptheta}), {}^{1}\bm{\uptheta}_1\right), 
\right. \\ & \; \left.
k_{\mathcal{D}_1}\left(z^\mathrm{e}_{29}({}^{2}\bm{\uptheta}),{}^{1}\bm{\uptheta}_1\right),\ldots,k_{\mathcal{D}_1}\left(z^\mathrm{e}_{40}({}^{2}\bm{\uptheta}),{}^{1}\bm{\uptheta}_1 \right)\right]^{\top},
\end{aligned}\\
\mathbf{k}_{\mathcal{D}_2}\left( \bm{\uptheta} \right)= \left[k_{\mathcal{D}_2}\left(z^\mathrm{e}_{13}({}^{2}\bm{\uptheta}), {}^{1}\bm{\uptheta}_2\right),\ldots,k_{\mathcal{D}_2}\left(z^\mathrm{e}_{28}({}^{2}\bm{\uptheta}), {}^{1}\bm{\uptheta}_2 \right)\right]^{\top}.
\end{gathered}
\end{equation}
Here, the random vectors $\mathbf{k}_{\mathcal{D}_1} \in \mathbb{R}^{24}$ and $\mathbf{k}_{\mathcal{D}_2} \in \mathbb{R}^{16}$ are parameterized by the lower-dimensional random vectors 
$^{1}\bm{\uptheta}_{1}\in\mathbb{R}^{M_{1,1}}$ and $^{1}\bm{\uptheta}_{2}\in\mathbb{R}^{M_{1,2}}$, respectively, through the K-L expansions. 
In summary, the target parameter vector components are $^{1}\bm{\uptheta}=\left[^{1}\bm{\uptheta}_{1}, {}^{1}\bm{\uptheta}_{2}\right]^{\top} \in \mathbb {R}^{M_{1,1} + M_{1,2}} $ and 
$^{2}\bm{\uptheta}=\left[z_{\mathrm{upper}},z_{\mathrm{lower}}\right]^{\top} \in \mathbb {R}^{M_2}$. The prior of $^{1}\bm{\uptheta}$ is $\mathcal{N}\left(\mathbf{0},\mathbf{I}_{M_{1,1} + M_{1,2}}\right)$, and the prior of $^{2}\bm{\uptheta}$ is set as a normal distribution $\mathcal{N}\left([2.5,7.5]^{\top}, \mathbf{I}_{2}\right)$ 
truncated between the constraints mentioned above. The parameters $\overline{u}$, $k_{\mathrm{min}}$, $v$, and the condition to determine the number of K-L expansion terms ($M_{1,1}$ and $M_{1,2}$) are the same as in the spatial-field-only case. The prior correlation lengths chosen for inversion are $l=2.5,5.0,10.0\,\mathrm{m}$. Here, it is expected that larger correlation lengths (relative to the spatial-field-only estimation case) should be sufficient to characterize the homogeneous spatial random fields as the geometry parameters represent the abrupt spatial changes in hydraulic conductivity.
In the proposed method, the bounding domain $\mathcal{D}'$ is defined as $\{z:0\leq z \leq10\}$, where the IEVP is solved with $N=25$ integration points (see \cref{fig:mesh1D}(C)) using the Nyström method. 

\begin{figure}[t]
\centering
\includegraphics{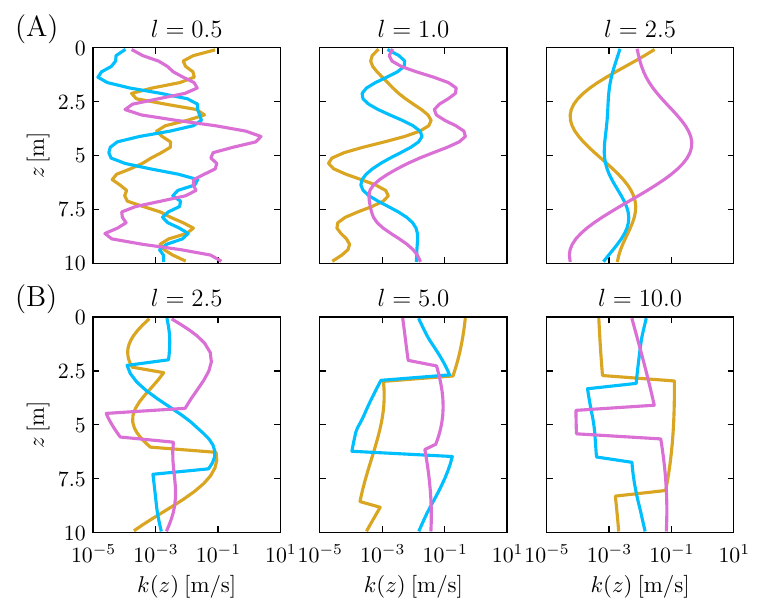}
\caption{\label{fig:prior1D} Hydraulic conductivity distributions corresponding to the realizations from the prior for (A) the spatial-field-only estimation case and (B) the simultaneous estimation case. The three color lines in each figure are three different realizations.}
\end{figure}

Realizations from the prior for different correlation lengths in the spatial-field-only estimation case and the simultaneous estimation case are shown in \cref{fig:prior1D}(A) and \cref{fig:prior1D}(B), respectively. In \cref{fig:prior1D}(A), the smaller $l$ is, the higher the frequency of spatial change of realizations, which indicates that a sufficiently small correlation length is necessary to capture rapid changes. However, sharp jumps cannot be achieved with such smooth priors. Incorporation of apriori information about the presence of a thin clay seam sandwiched between sandy layers, through geometry parameters, enables sampling of hydraulic conductivity profiles (see \cref{fig:prior1D}(B)) that display spatial jumps.

Five Markov chains are generated from different starting points sampled from the prior distribution, each containing 25000 samples. Automatic parameter tuning of the leapfrog parameters $\varepsilon$ and $L$, and the mass matrix $\bm{\mathcal{M}}$ (in a manner similar to Stan's HMC adaptation \cite{carpenter2017}) is done during the first 5000 iterations, consisting of the burn-in period.
The initial step-size is set to $\varepsilon=10^{-6}$, and the mass matrix is chosen as the inverse of the covariance matrix of the prior of $\bm{\uptheta}$, i.e., $\bm{\mathcal{M}} = \bm{\Sigma}_{\bm{\uptheta}}^{-1}$ (see \cref{eq:prior}).

\begin{figure}
\centering
\includegraphics[width=0.96\textwidth]{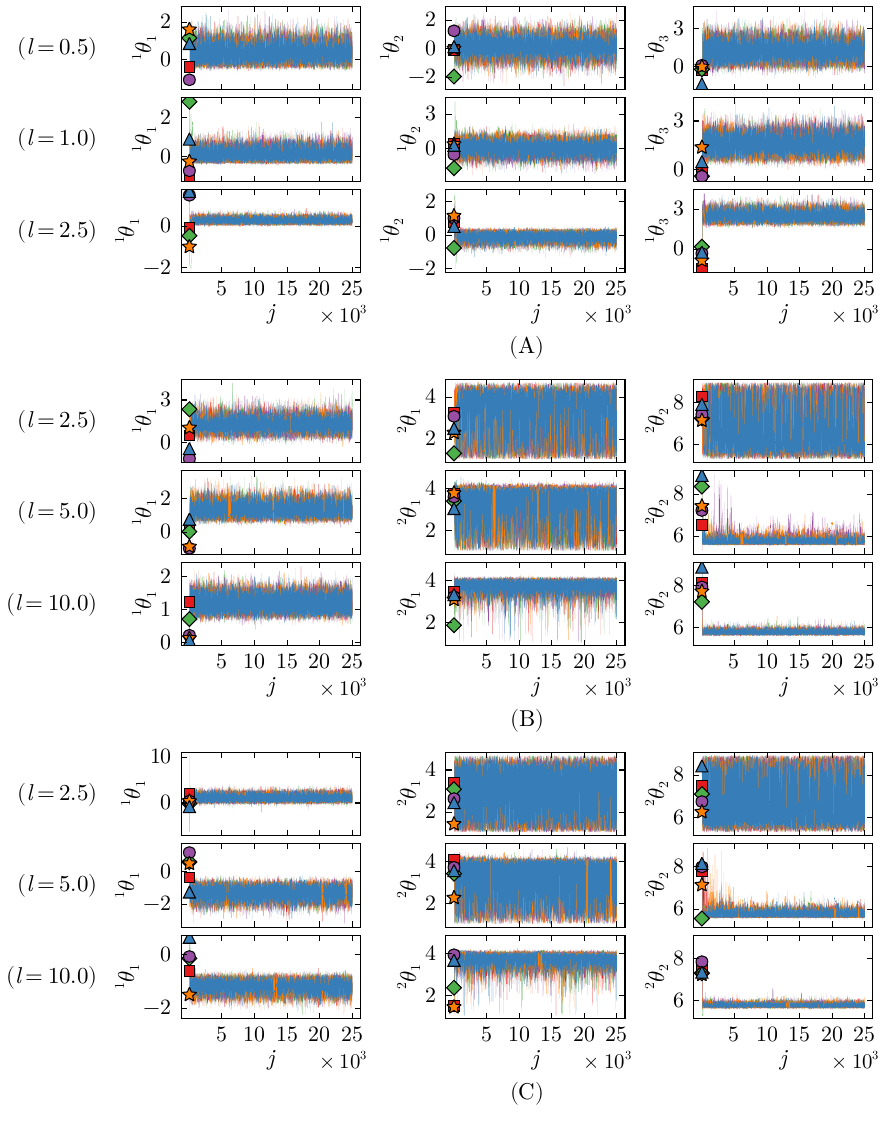}
\caption{\label{fig:markov_chains1D} Trace plots of Markov chains for (A) $^1\theta_1$, $^1\theta_2$, and $^1\theta_3$ for the spatial-field-only estimation method, (B) $^1\theta_1$, $^2\theta_1$, and $^2\theta_2$ for the simultaneous estimation method of Koch et al. (2021) \cite{koch2021}, and (C) $^1\theta_1$, $^2\theta_1$, and $^2\theta_2$ for the proposed simultaneous estimation method. The five color lines represent different chains, and the five markers represent different initial values.
}
\end{figure}

Samples from the burn-in period are discarded during posterior inference. Trace plots of the Markov chains are shown in \cref{fig:markov_chains1D}, indicating all the chains with different starting points converge in the same range. Convergence of Markov chains is also confirmed by the multivariate effective sample size ($\mathrm{mESS}$) \cite{vats2019, vats2021, roy2020}. The Markov chain is considered convergent when the consistent estimator of $\mathrm{mESS}$, which is obtained from the generated samples, is above a threshold value (called minimum ESS) as follows:
\begin{equation}
\widehat{\mathrm{mESS}}\geq\frac{2^{2/M}\pi}{\left(M\Gamma\left(M/2\right)\right)^{2/M}}\frac{\chi_{1-\alpha_{\mathrm{ESS}}, M}^{2}}{\epsilon_{\mathrm{ESS}}^{2}},
\label{eq:multiESS}
\end{equation}
where the RHS is the minimum ESS, $\epsilon_{\mathrm{ESS}}$ is the desired level of precision for the volume of $100(1-\alpha_{\mathrm{ESS}})\%$ asymptotic confidence interval, $\chi_{1-\alpha_{\mathrm{ESS}}, M}^{2}$ is the $1 - \alpha_{\mathrm{ESS}}$ quantile of the chi-squared distribution $\chi_M^2$, and $\Gamma$ is the gamma function. \cref{tab:ESS1D} shows $M$ (the dimension of $\bm{\uptheta}$), the mESS, and the minimum ESS for each analysis. All mESS values are larger than each minimum ESS, indicating that the samples are well converged to the target distributions.
Here, it must be noted that, because the simultaneous estimation case allows for the use of relatively larger correlation lengths, the dimensionality of the inverse problem $M$ (see \cref{tab:ESS1D}) can be reduced in comparison to the spatial-field-only case. This points to the fact that the introduction of geometry parameters does not complicate the analysis, but rather simplifies it and helps alleviate the curse of dimensionality.

\begin{table}[t]
\caption{Minimum ESS and multivariate ESS for the three inversion methods. The number of K-L expansion terms chosen is related to the criterion $\lambda_{M_1}>10^{-3}$.}
\begin{center}
\begin{tabular}{ccccc}
\hline 
 & $l$ & $M$ & minESS & mESS\tabularnewline
\hline 
\multirow{3}{*}{Spatial-field-only \cite{koch2020a}} & 0.5 & 28 & 8592 & 26906.66 \tabularnewline
 & 1.0 & 15 & 8793 & 29189.73 \tabularnewline
 & 2.5 & 8 & 8804 & 38391.09 \tabularnewline
\hline 
\multirow{3}{*}{Simultaneous \cite{koch2021}} & 2.5 & 13 & 8817 & 16918.24 \tabularnewline
 & 5.0 & 10 & 8831 & 10135.66 \tabularnewline
 & 10.0 & 9 & 8823 & 21057.63 \tabularnewline
\hline 
\multirow{3}{*}{Simultaneous (proposed)} & 2.5 & 16 & 8778 & 38176.65 \tabularnewline
 & 5.0 & 12 & 8826 & 12946.18 \tabularnewline
 & 10.0 & 8 & 8804 & 15959.50 \tabularnewline
\hline 
\end{tabular}
\end{center}
\label{tab:ESS1D}
\end{table}

\begin{figure}[t!]
\centering
\includegraphics{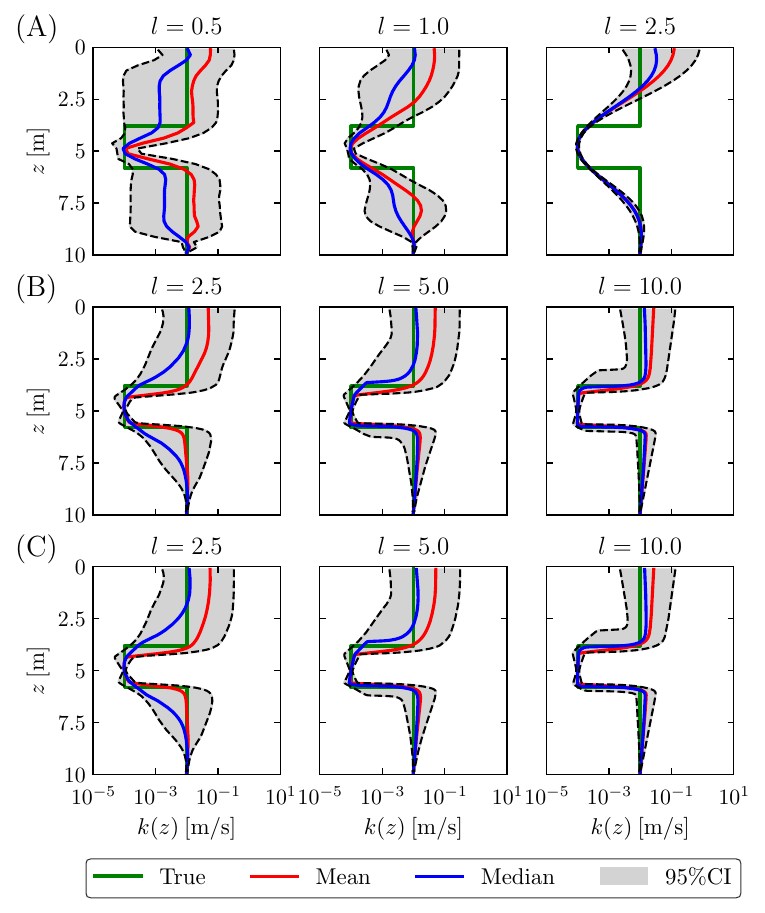}
\caption{\label{fig:Estimated-fields}Hydraulic conductivity fields obtained
by (A) the spatial-field-only estimation, (B) the simultaneous estimation method
of Koch et al. (2021) \cite{koch2021}, and (C) the proposed simultaneous estimation method.}
\end{figure}

Results from the spatial-field-only estimation method for three different prior correlation lengths are shown in \cref{fig:Estimated-fields}(A). In the case of $l=2.5\,\mathrm{m}$, the hydraulic conductivity tends to be small in the central part of the domain. However, the 95\% CI does not envelop the true spatial field. This is because $l$ is large relative to the width of the central part of the domain with different hydraulic conductivity, and the K-L expansion is unable to resolve abrupt spatial changes in hydraulic conductivity. In the case of $l=1.0\,\mathrm{m}$, the 95\% CI is larger than in the case of $l=2.5\,\mathrm{m}$, but it fails to envelop the true spatial field and also does not capture the spatial change in hydraulic conductivity. In the case of $l=0.5\,\mathrm{m}$, the 95\% CI envelops most of the true spatial field. However, this result is not sufficiently accurate, as there is a large variance around both ends of the domain. 

\begin{figure}[bp!]
\centering
\includegraphics[height=0.80\textheight]{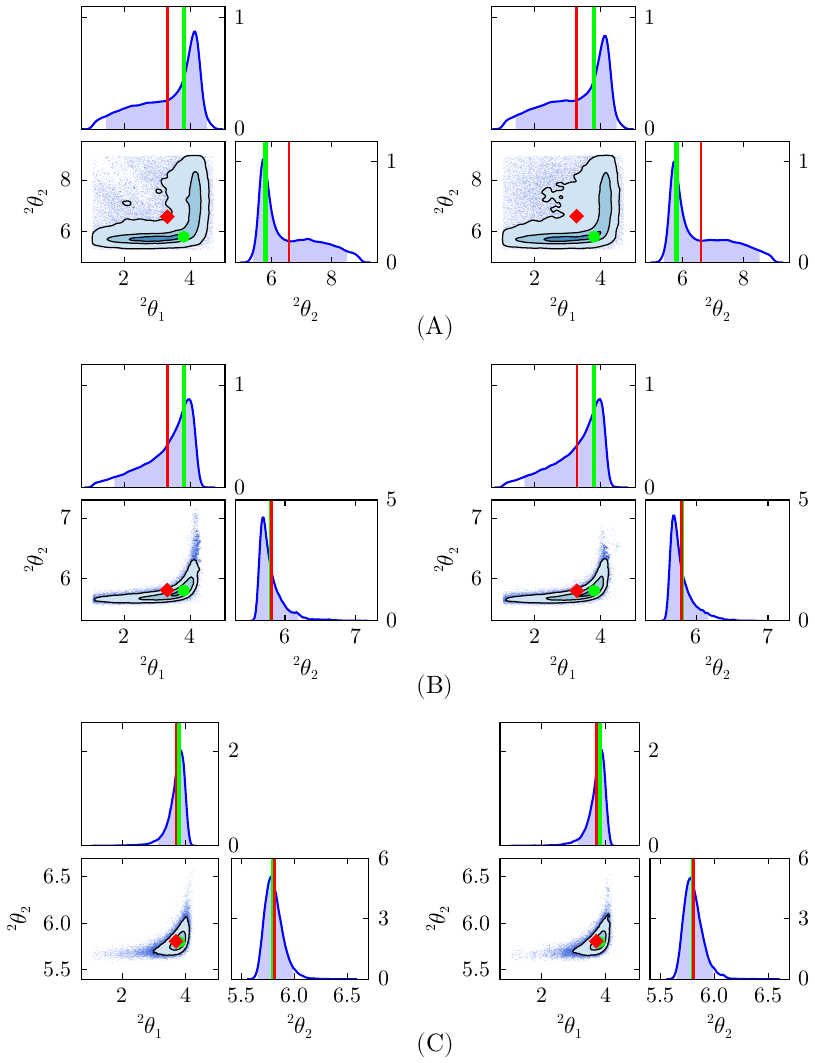}
\caption{\label{fig:Pairplots1D}Marginal distributions of $p\left(^{2}\bm{\uptheta}|\mathbf{y}\right)$
obtained by the simultaneous estimation methods for (A) $l=2.5\,\mathrm{m}$, (B) $l=5.0\,\mathrm{m}$, and (C) $l=10.0\,\mathrm{m}$. In each case (A), (B), and (C), the two left-hand columns correspond to the results of the method of Koch et al. (2021) \cite{koch2021}, and the two right-hand columns correspond to the results of the proposed method. The green circle and green line represent the true value of $^{2}\bm{\uptheta}$. The red diamond and red line represent the mean value of the posterior distribution. In 1D distribution plots, the blue color area represents 95\% High Probability Density (HPD) regions, and in 2D distribution plots, the three contours indicate the 5\%, 50\%, and 95\% HPD regions.}
\end{figure}

The estimated hydraulic conductivity spatial field using the simultaneous estimation method of Koch et al. (2021) \cite{koch2021} and the proposed simultaneous estimation method are shown in \cref{fig:Estimated-fields}(B) and \cref{fig:Estimated-fields}(C), respectively.
The results of both methods are almost identical. In other words, the proposed method does not compromise the performance of the simultaneous estimation of Koch et al. (2021) \cite{koch2021}, and the small error due to the truncation of the K-L expansion associated with the domain independence property (see \cref{subsec:DomainIndep}) has a negligible effect on the inversion results.
For all correlation lengths, the 95\% CIs envelop most of the true spatial fields. In particular, in the case of $l=10.0\,\mathrm{m}$, the incorporation of geometry parameters $^{2}\bm{\uptheta}$ clearly enables HMC to resolve jumps in the spatial field, as seen in the posterior mean, median, and narrow 95\% CI. However, the width of the 95\% CIs increases, and the resolution of the jump decreases as the prior correlation length decreases. This difference due to the correlation length can be explained as follows. Spatial field realizations with smaller correlation lengths display a higher frequency of spatial oscillation (see \cref{fig:prior1D}(B)) in $\mathcal{D}_1$ and $\mathcal{D}_2$ around the true (constant) hydraulic conductivity field. Among these realizations, there can be many spatial fields, which while differing from the true spatial conductivity field, generate a forward response such that the error from the observation data is nearly identical. This results in a situation where the posterior has a large variance and is influenced by the prior relatively more than by the likelihood. This trend can also be seen in the posterior distributions of ${}^2\bm{\uptheta}$ in \cref{fig:Pairplots1D}. The effect of the prior in the posterior is more significant at smaller correlation lengths, as seen in \cref{fig:Pairplots1D}(A), wherein a larger number of samples are skewed towards the normal prior $^{2}\bm{\uptheta} \sim \mathcal{N}\left([2.5,7.5]^{\top}, \mathbf{I}_{2}\right)$ truncated between constraints mentioned earlier in this section. This behavior leads to a larger variance and an inability to differentiate between geometry parameters in the posterior, and the mean and median of the hydraulic conductivity spatial field appear similar to the case in which geometry parameters are not considered. Consequently, these results highlight the importance of an appropriate choice of correlation length parameters in the inverse analysis.

%% ==================================================================
\subsubsection{Comparison of computation times for the two simultaneous estimation methods}\label{subsubsec:1Dtime}
In this section, the computation time for the derivative of the K-L expansion $\frac{\partial \mathbf{u}}{\partial \bm{\uptheta}}$ and one leapfrog step are measured for the two simultaneous estimation methods. To be precise, the total computation time for $\varphi$ and $\frac{\partial \varphi}{\partial \bm{\uptheta}}$, which accounts for the majority of the leapfrog computation time, is measured and considered as the computation time of one leapfrog step. The number of terms in the K-L expansion is $M_1 = 10$ for both cases, and the correlation length is $l=10.0\,\mathrm{m}$. Different meshes with 40, 120, 240, 400, 640, 1000, 1600, 2520, 4000, 6400, and 10000 finite elements $n_\mathrm{e}$ are considered, and $\bm{\uptheta}$ is set to $^{1}\bm{\uptheta}=\bm{0}$ and $^{2}\bm{\uptheta}=[3,7]^\top$ as representative values. The parameters $\overline{u}$, $v$, the BCs, the constraints of $^2 \bm{\uptheta}$, the prior of $\bm{\uptheta}$, and the observation data are the same as in \cref{subsubsec:1Dresult}. The HMC parameters $\varepsilon$ and $\bm{\mathcal{M}}$ are set to the same initial values as in \cref{subsubsec:1Dresult}.

\begin{figure}[!t]
\centering
\includegraphics{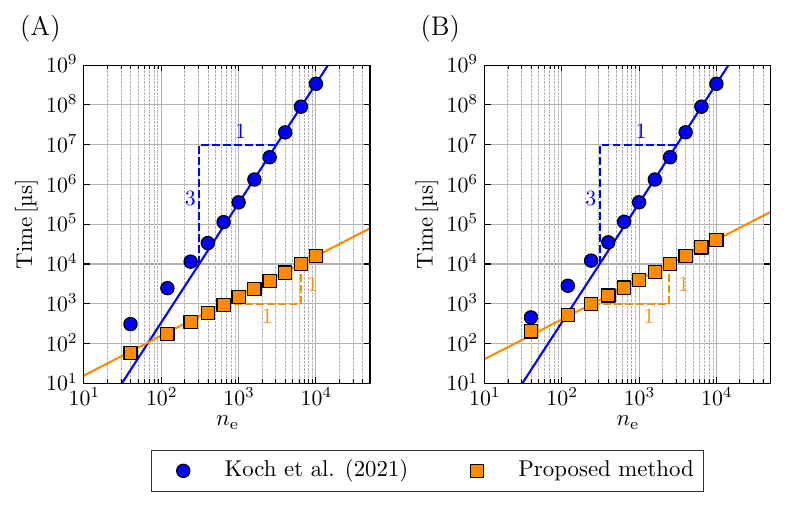}
\caption{\label{fig:CompTime} 
Computation time of (A) the derivative of the K-L expansion and (B) one leapfrog step.
}
\end{figure}

The computational time of the derivative of the K-L expansion for the two estimation methods is shown in \cref{fig:CompTime}(A). The slope for the method of Koch et al. (2021) \cite{koch2021} asymptotically approaches 3, demonstrating that the time complexity is $O(n_\mathrm{e}^{3})$. This result is due to the Moore-Penrose inverse, whose time complexity is $O(n_\mathrm{e}^3)$, which is derived from the singular value decomposition \cite{demmel1997}. On the other hand, the slope for the proposed method is approximately 1, indicating that the time complexity order is $O(n_\mathrm{e})$. This is because the derivative of the eigenfunction, which has time complexity independent of $n_\mathrm{e}$ (see \cref{eq:dPhi_d2theta}), is computed for $n_\mathrm{e}$ positions. Therefore, the time complexity can be improved by approximately two orders of magnitude through the use of the domain independence property, which avoids the computation of the Moore-Penrose inverse.

\cref{fig:CompTime}(B) shows the computation time for one leapfrog step, which includes the cost of the finite element forward solver. The slope of the line for the method of Koch et al. (2021) \cite{koch2021} is almost the same as in \cref{fig:CompTime}(A), indicating that the main source of the computational cost in the simultaneous method of Koch et al. (2021) \cite{koch2021} is the derivative of the K-L expansion. On the other hand, the line for the proposed method is shifted up compared to \cref{fig:CompTime}(A) but retains a slope of 1. This implies that the computation time of the derivative of the K-L expansion is not dominant in the entire analysis. 
The unit slope refers to an $O(n_\mathrm{e})$ time complexity, which is to be expected. This is because the computational cost of the forward analysis is also $O(n_\mathrm{e})$ since the global matrix in this 1D seepage problem is symmetric and tridiagonal.
In more general cases in 2D and 3D, the time complexity of forward analysis, using FEM, ranges from $O(n_\mathrm{e})$ to $O(n_\mathrm{e}^3)$, where the exact time complexity depends on the linear solver and the sparsity of the global matrix \cite{demmel1997, george1973}. Therefore, in the simultaneous method of Koch et al. (2021) \cite{koch2021}, the computation of the derivative of the K-L expansion is the primary source of the computational cost. In contrast, this computation is avoided and is no longer a bottleneck in the proposed method.

%% ==================================================================
\subsection{2D seepage flow problem}

\begin{figure}[!t]
\centering
\includegraphics[width=1.0\textwidth]{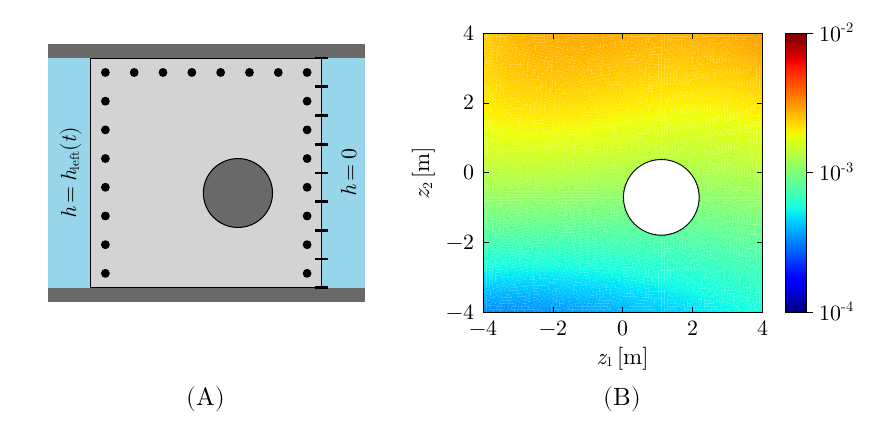}
\caption{\label{fig:Setup-2D}(A) Target domain containing a circular cavity with an impermeable boundary. Observations include hydraulic heads at 22 black circle points and outflow discharge rate from 8 sections on the right edge. (B) True hydraulic conductivity field generated by the K-L expansion with $l=8\,\mathrm{m}$.}
\end{figure}

The second inverse problem shown in \cref{fig:Setup-2D} is related to 2D seepage flow. The target domain has a circular cavity with an impermeable boundary. The true center coordinates $(z_1^{\mathrm{center}},z_2^{\mathrm{center}})$ and radius $r$ of the circular cavity are $(z_1^{\mathrm{center}},z_2^{\mathrm{center}}, r)=(1.1,-0.7,1.1)$, which are considered unknown in the inverse problem. The true spatial field of hydraulic conductivity is generated numerically through ${u}$ expressed by the K-L expansion with 
%correlation length $l=8\,\mathrm{m}$, 
parameters $(v, l, \overline{u}) = (1, 8, -3)$, and is shown in \cref{fig:Setup-2D}(B).
Dirichlet BCs are implemented such that the hydraulic head on the left end varies as $h_\mathrm{left}(t) = 0.1 + 0.01(t-1)$ for $ t = \{1,...,31\}$, while the hydraulic head on the right end is fixed as $h = 0$. Zero water flux Neumann BCs are implemented on the top and bottom sides of the domain and on the circular cavity boundary. Observation data consists of hydraulic heads at 22 points (black circles in \cref{fig:Setup-2D}(A)) and outflow discharge rate from 8 sections (divided by black horizontal bars in \cref{fig:Setup-2D}(A)). The data is generated numerically by adding zero mean Gaussian noise with standard deviation (10\% of the true value) to the true values, obtained by solving the forward problem using FEM. 
\begin{figure}[!t]
\centering
\includegraphics[width=1.0\textwidth]{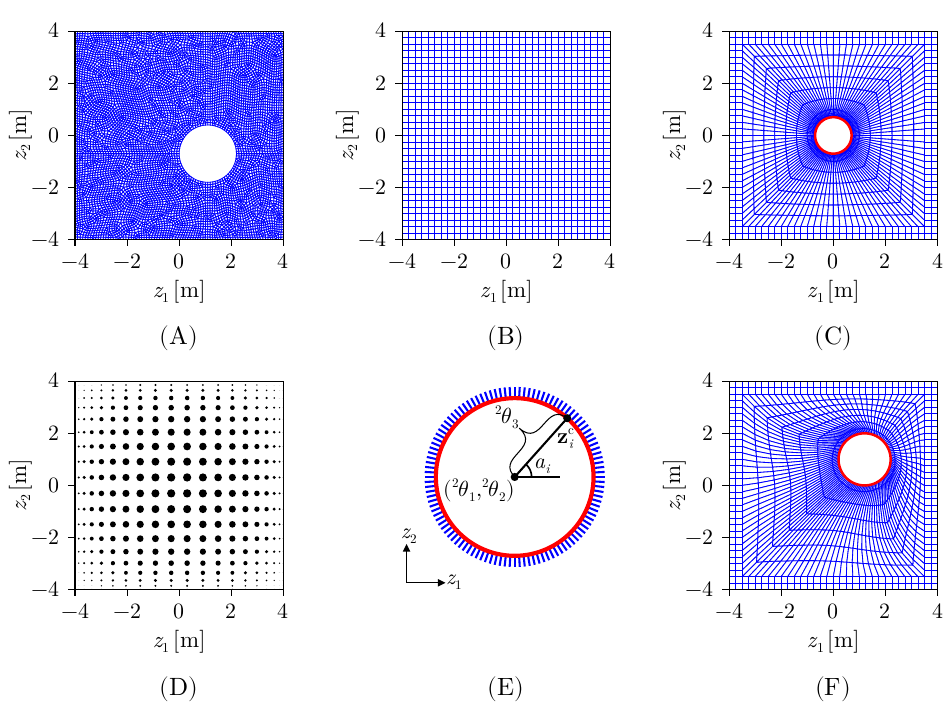}
\caption{\label{fig:Mesh2D}
(A) Mesh for generating observation data.
(B) Mesh for the spatial-field-only estimation.
(C) Reference mesh used in the simultaneous estimation method. 
(D) Gauss quadrature points on the bounding domain $\mathcal{D}'$.
(E) Parameterization of the cavity boundary.
(F) Mesh for $\mathcal{D}(^{2}\bm{\uptheta})$ 
obtained by applying enforced displacements 
dependent on $^{2}\bm{\uptheta}$ to the reference mesh.
}
\end{figure}

The spatial-field-only estimation does not assume the presence of the water-impermeable zone inside the target domain and is performed on a uniform structured mesh 
with elements 
of size \mbox{0.25 m}, which is coarser than the observed mesh (see \cref{fig:Mesh2D}(A)). The target to be estimated is the 1024-dimensional hydraulic conductivity random vector
$\mathbf{k}\left( {}^1\bm{\uptheta} \right) =\left[k\left(\mathbf{z}_1^\mathrm{e}, {}^1\bm{\uptheta} \right), \ldots k\left(\mathbf{z}_{1024}^\mathrm{e}, {}^1\bm{\uptheta}\right)\right]^{\top}$, 
whose entries correspond to 1024 finite elements in \cref{fig:Mesh2D}(B).
The random vector $^{1}\bm{\uptheta}\in\mathbb{R}^{M_1}$ is derived from the $M_1$-term truncated K-L expansion of $u$ with the mean $\overline{u}=-3$, and its prior is $\mathcal{N}\left(\mathbf{0},\mathbf{I}_{M_1}\right)$.
The criterion for selecting $M_1$ is $\lambda_{M_1}>\lambda_{\mathrm{max}}\times10^{-3}$,
where $\lambda_{\mathrm{max}}$ is the maximum eigenvalue, and the lower bound of $k$ is set $k_{\mathrm{min}}=10^{-7} \, \mathrm{m/s}$. The scale of the autocovariance matrix is set to $v=1$, and two prior correlation lengths, i.e., $l=2,4\,\mathrm{m}$, are chosen to study the inversion results. 

Simultaneous spatial field and geometry estimation is performed only by the proposed method. This estimation assumes the presence of the water-impermeable zone inside the target domain. Therefore, the targets to be estimated are the geometry parameter vector $^{2}\bm{\uptheta}=\left[z_1^{\mathrm{center}},z_2^{\mathrm{center}}, r\right]^{\top}$ and the 1024-dimensional hydraulic conductivity random vector 
$\mathbf{k}\left(\bm{\uptheta} \right)= \left[k\left(\mathbf{z}_1^\mathrm{e}({}^2\bm{\uptheta}), {}^1\bm{\uptheta} \right),\ldots,k\left(\mathbf{z}_{1024}^\mathrm{e}({}^2\bm{\uptheta}), {}^1\bm{\uptheta}\right)\right]^{\top}$, whose entries correspond to 1024 finite elements in \cref{fig:Mesh2D}(C). 
The random vector ${}^1\bm{\uptheta}$ is derived from the $M_1$-term truncated K-L expansion of $u$.
The eigenvalues and eigenfunctions in the K-L expansion are obtained by solving the IEVP on the bounding domain $\mathcal{D}'$ defined as $\{(z_1,z_2):-4\leq z_1 \leq 4,$ $-4 \leq z_2 \leq 4$\}.
The $20\times20$ grid of Gauss quadrature points $(N=400)$ for numerical integration of the IEVP are shown in \cref{fig:Mesh2D}(D). 
%The prior of $^{1}\bm{\uptheta}$
The parameters $\overline{u}$, $k_{\mathrm{min}}$, $v$, and the condition to determine $M_1$ are set identically to those in the spatial-field-only estimation case. Inversion is performed considering two correlation lengths, i.e., $l=4,8\,\mathrm{m}$. Additionally, through the introduction of geometry parameters, it is possible to accurately implement the impermeable boundary conditions on the circular cavity boundary.  

The mesh in \cref{fig:Mesh2D}(C) moves each time $^{2}\bm{\uptheta}$ is updated according to the reversible mesh moving method detailed in Koch et al. (2020) \cite{koch2020b}. This method implements mesh movement by elastically deforming a fixed reference mesh, which is the starting point of the deformation, through prescribed displacements. Since all deformation takes place through the reference mesh, the deformation is guaranteed to be reversible. Updates in $^{2}\bm{\uptheta}$ define the new position of the cavity boundary, thereby yielding the prescribed displacements of the cavity boundary nodes. In this analysis, the updated position of the 112 nodes describing the boundary of the circular cavity is given as

\begin{equation}
\mathbf{z}_{i}^{\mathrm{c}}({}^{2}\bm{\uptheta})=\left(\begin{array}{c}
^{2}\theta_{1}+{}^{2}\theta_{3}\cos{a_i}\\
^{2}\theta_{2}+{}^{2}\theta_{3}\sin{a_i}
\end{array}\right),\quad{a_i}=\frac{2\pi}{112}(i-1),\label{eq:coords_on_circle}
\end{equation}
where $\mathbf{z}_{i}^{\mathrm{c}}$ refers to the coordinates of $i$th node on the circle, $a_{i}$ is the angle corresponding to $i$th node as shown in \cref{fig:Mesh2D}(E). 
During this mesh moving phase, the BC is set to zero displacement on the outer boundaries of the domain, and the observation points are kept fixed. A decent mesh quality (see \cref{fig:Mesh2D}(F)) is maintained by considering an artificial elastic modulus that is scaled by the determinant of the element Jacobian. The elemental stiffness matrix on the reference mesh is written as
\begin{equation}
\mathbf{K}_{e}^{\mathrm{ref}}=\int_{\mathcal{\mathcal{D}}_{e}}\mathbf{B}^{\mathrm{ref}\top}\mathbf{D}^{\mathrm{ref}}\mathbf{B}^{\mathrm{ref}}\left|\mathbf{J}_{e}^{\mathrm{ref}}\right|\left(\frac{J^0}{\left|\mathbf{J}_{e}^{\mathrm{ref}}\right|}\right)^{\chi}\mathrm{d}\mathcal{D}_{e},
\end{equation}
where $\mathbf{J}_{e}^{\mathrm{ref}}$ is Jacobian, $\mathbf{B}^{\mathrm{ref}}$ is a strain-displacement matrix, $\mathbf{D}^{\mathrm{ref}}$ is the constitutive matrix, $\chi>0$ is stiffening power, and $J^0$ is an arbitrary scaling parameter.
For all elements, the Young's modulus and Poisson's ratio, which determine
$\mathbf{D}^{\mathrm{ref}}$, are set at 2500 MPa and 0.25, respectively, while both parameters $J^0$ and $\chi$ are set to 1.
The prior of $^{2}\bm{\uptheta}$ is a truncated Gaussian defined through the Gaussian density $\mathcal{N}\left([0,0,0.5]^{\top},\mathbf{I}_{3}\right)$ and the constraints:
\begin{equation}
\begin{gathered}\lvert{}^{2}\theta_{1}\rvert<2.1-\frac{^{2}\theta_{3}}{2}\\
\lvert{}^{2}\theta_{2}\rvert<2.1-\frac{^{2}\theta_{3}}{2}\\
0.1<^{2}\theta_{3}<1.6
\end{gathered}
\end{equation}
to avoid mesh breakage and non-physical realizations of the circular cavity.

\begin{figure}
\centering
\includegraphics[height=0.86\textheight]{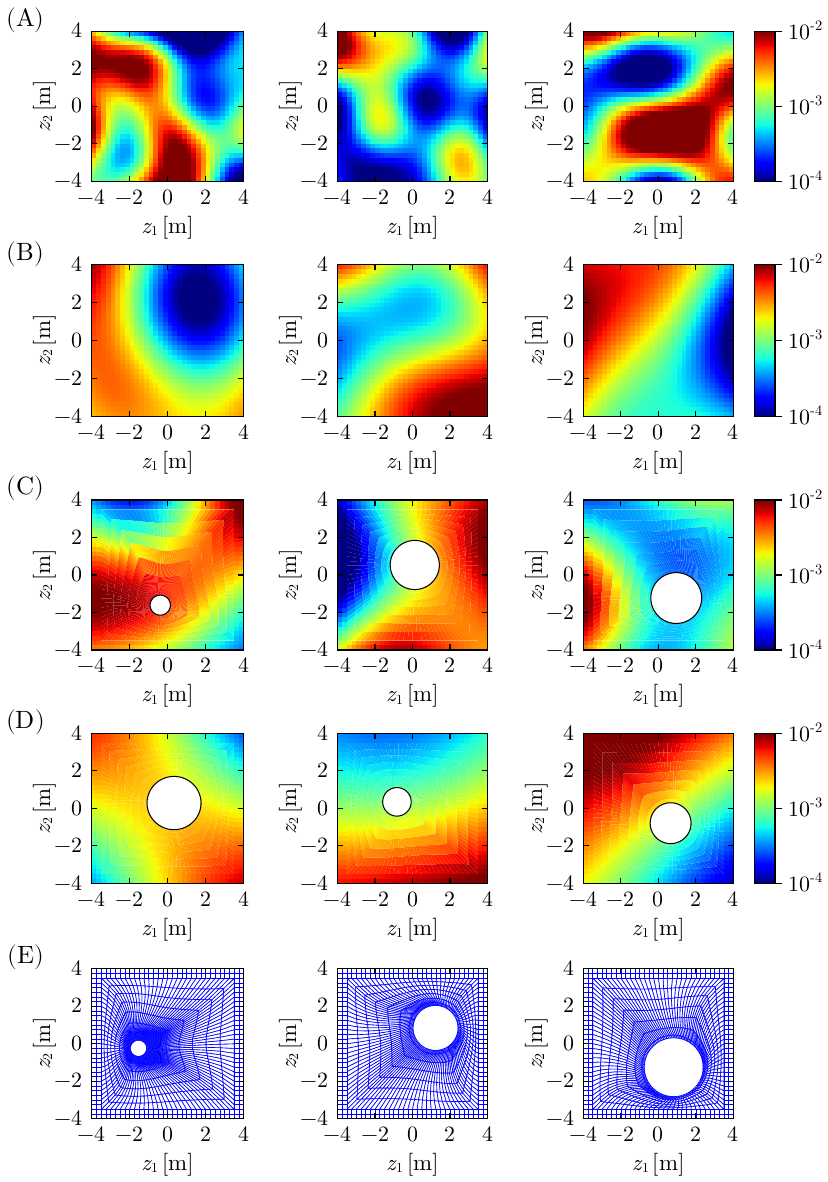}
\caption{\label{fig:prior2D} Realizations from the prior for the spatial-field-only estimation method with (A) $l=2\,\mathrm{m}$ and (B) $l=4\,\mathrm{m}$, and for the simultaneous estimation method with (C) $l=4\,\mathrm{m}$ and (D) $l=8\,\mathrm{m}$. Meshes corresponding to the realizations from the prior of the geometry parameter $^{2}\bm{\uptheta}$ are shown in (E). Maintenance of a good mesh quality is confirmed in the realizations.}
\end{figure}

Realizations from the prior for the two estimation approaches are shown in \cref{fig:prior2D}. In the spatial-field-only estimation method, as in the 1D case, it is obvious that small correlation lengths $l$ are necessary to capture rapidly oscillating spatial fields. Additionally, it is clear from \cref{fig:prior2D}(A) and \cref{fig:prior2D}(B), that the spatial change of the hydraulic conductivity field is gradual, suggesting that even if the rough location of the water-impermeable circle is detected, the location of its interface would not be estimated clearly. In contrast, in \cref{fig:prior2D}(C) and \cref{fig:prior2D}(D), the realizations include holes due to the introduction of geometry parameters. This explicitly provides interface locations in the simultaneous estimation results. For completeness, \cref{fig:prior2D}(E) shows meshes corresponding to realizations from the prior of geometry parameters in the simultaneous estimation method.

Four chains, each containing 12,000 samples, are obtained using HMC, starting from points randomly drawn from the prior distribution.
Similar to the 1D analysis, automatic parameter tuning of the leapfrog parameters $\varepsilon, L$, and the mass matrix $\bm{\mathcal{M}}$ is done during the first 2000 iterations. Samples from this burn-in period are discarded during posterior inference.

\begin{figure}[!t]
\centering
\includegraphics[width=1.0\textwidth]{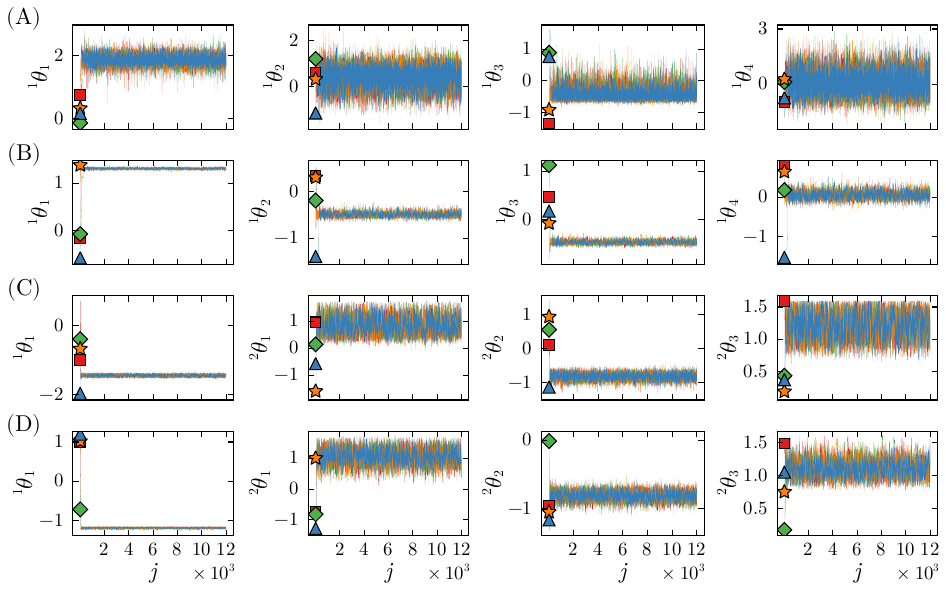}
\caption{\label{fig:markov_chains2D} Trace plots of Markov chains. Each row corresponds to a different analysis. The top two rows represent chains of $^1\theta_1$, $^1\theta_2$, $^1\theta_3$, and $^1\theta_4$ for the spatial-field-only estimations with (A) $l=2\,\mathrm{m}$ and (B) $l=4 \, \mathrm{m}$. The bottom two rows represent chains of $^1\theta_1$, $^2\theta_1$, $^2\theta_2$, and $^2\theta_3$ for the proposed simultaneous estimation method with (C) $l=4\,\mathrm{m}$ and (D) $l=8\,\mathrm{m}$. The four color lines represent different chains, and the four markers represent different initial values.}
\end{figure}

As in the 1D case, the Markov chain convergence is confirmed by trace plots and the mESS. In all trace plots (see \cref{fig:markov_chains2D}), the chains originating from different starting points converge in the same range, and each chain converges to its high probability region quickly. In \cref{tab:ESS2D}, all mESS values are larger than each minimum ESS. Based on these two criteria, we judge the Markov chains to be converged. Once again, fewer terms $M$ are required in the simultaneous estimation case, as the discontinuities in the spatial field are captured by geometry parameters explicitly. 

%=======================================================================================
%Preferably on the same page.

\begin{table}[!b]
\caption{Minimum ESS and multivariate ESS.}
\begin{center}
\begin{tabular}{ccccc}
\hline 
 & $l$ & $M$ & minESS & mESS \tabularnewline
\hline 
\multirow{2}{*}{Spatial-field-only \cite{koch2020a}} & 2 & 28 & 8592 & 26036.70 \tabularnewline
 & 4 & 11 & 8831 & 10535.25 \tabularnewline
\hline 
\multirow{2}{*}{Simultaneous (proposed)} & 4 & 14 & 8806 & 10398.88 \tabularnewline
 & 8 & 9 & 8823 & 11851.81 \tabularnewline
\hline 
\end{tabular}
\end{center}
\label{tab:ESS2D}
\end{table}

%=======================================================================================
\begin{figure}
\centering
\includegraphics[height=0.84\textheight]{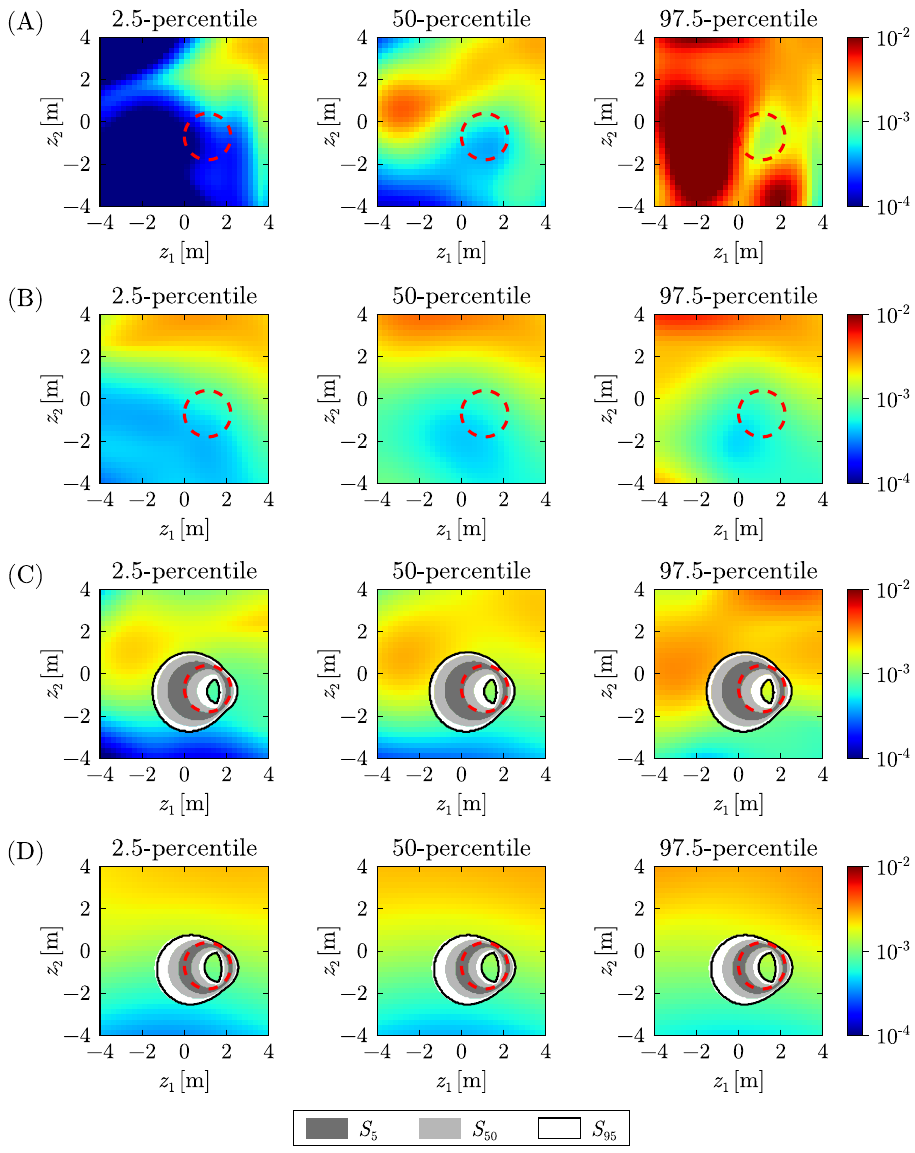}
\caption{\label{fig:hc2D} 
Hydraulic conductivity fields obtained by the spatial-field-only estimation method with (A) $l=2\,\mathrm{m}$ and (B) $l=4\,\mathrm{m}$, and by the simultaneous estimation method with (C) $l=4\,\mathrm{m}$ and (D) $l=8\,\mathrm{m}$. The 2.5- and 97.5-percentile plots correspond to the ends of 95\% CI. (C) and (D) also show contour plots indicating the uncertainty in the location of the boundary of the circular cavity. Darker colors represent higher probability that an interface is present. The red dotted circles show the true location of the interface.} 
\end{figure}

The results of the spatial-field-only estimation are shown in \cref{fig:hc2D}(A) and \cref{fig:hc2D}(B) with the 2.5, 50, and 97.5 percentiles of the hydraulic conductivity. In the case of $l=4\,\mathrm{m}$, the estimated spatial field is similar to the true one in that the hydraulic conductivity tends to be higher at the top and lower at the bottom. The zone with relatively low hydraulic conductivity in the lower right quadrant of the domain reflects the presence of an impermeable area. However, this low hydraulic conductivity zone deviates from the true position of the circle (corresponding to the red circle in \cref{fig:hc2D}). Additionally, it is almost impossible to demarcate the boundary of this impermeable zone. In the case of $l=2\,\mathrm{m}$, the 50-percentile result has a similar hydraulic conductivity trend to the true field, varying from larger values at the top to lower values at the bottom of the domain. However, the 2.5-percentile and the 97.5-percentile, the boundaries of 95\% CI, have more than 2 orders of magnitude differences. The large variance in the estimation results makes it difficult to draw useful conclusions of the uncertainty in the location of the impermeable circular cavity and the surrounding hydraulic conductivity spatial field.

The results of the proposed simultaneous estimation method are shown in \cref{fig:hc2D}(C) and \cref{fig:hc2D}(D). The heatmap of the hydraulic conductivity spatial field is overlain with a contour plot reflecting the location of the impermeable boundary of the circular cavity. 
In both cases of $l=4\,\mathrm{m}$ and $8\,\mathrm{m}$, the estimated spatial fields are similar to the true field. The 95\% CIs are not large and envelope the true fields in most locations, indicating the high accuracy of the estimation results with respect to the spatial field.
The grey-scale contour plot has three zones with different color depths and two black contours. Darker colors represent a higher probability that an interface is present, and the zone bounded by the two black contours corresponds to the 95\% HPD (High Probability Density) of $p\left(^{2}\bm{\uptheta}|\mathbf{y}\right)$. In more detail, we define $S_{\alpha}$ as the smallest zone that completely encompasses all circle interfaces generated by samples belonging to the $\alpha$\% HPD region. Then, the zone within the darkest color is $S_{5}$, within the darkest and second darkest color is $S_{50}$, and within all three zones is $S_{95}$. Clearly, the 95\% HPD region (between the two black contours) envelops the impermeable boundary of the circular cavity in both cases $l=4\,\mathrm{m}$ and $8\,\mathrm{m}$. In particular, in the case of $l=8\,\mathrm{m}$, the darkest color zone matches the true interface well, indicating a high accuracy of the estimated results.

\begin{figure}
\centering
\includegraphics[height=0.84\textheight]{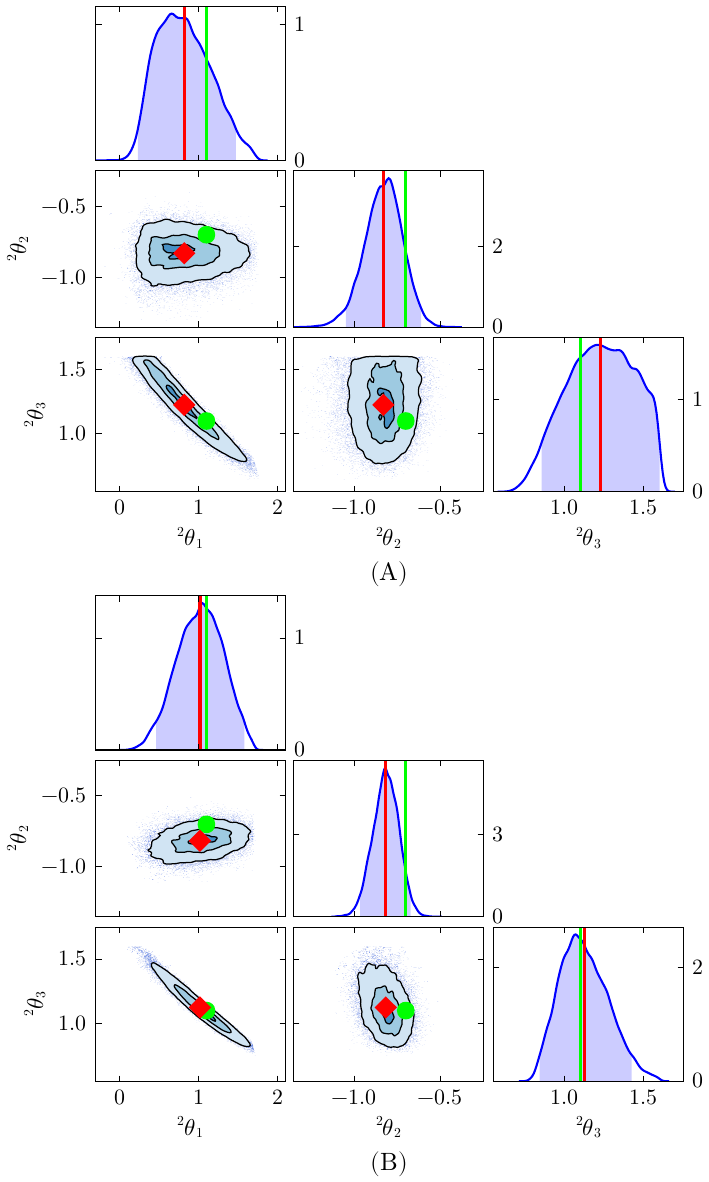}
\caption{\label{fig:Pairplots2D}Marginal distributions of $p\left(^{2}\bm{\uptheta}|\mathbf{y}\right)$
obtained by the simultaneous estimation method with (A) $l=4\,\mathrm{m}$ and (B)
$l=8\,\mathrm{m}$. The green circle and green line represent the true value of $^{2}\bm{\uptheta}$.
The red diamond and red line represent the mean value of the posterior distribution.
In 1D distribution plots, the blue color area represents 95\% HPD regions, and in 2D distribution plots, the three contours indicate the 5\%, 50\%, and 95\% HPD regions.}
\end{figure}

The marginal distributions of the posterior of the geometry parameters $^{2}\bm{\uptheta}$ in the simultaneous estimation case are shown in \cref{fig:Pairplots2D}. 
In the case of $l=4\,\mathrm{m}$ (\cref{fig:Pairplots2D}(A)), the 95\% HPDs of $p\left(^2\theta_1|\mathbf{y}\right)$, $p\left(^2\theta_2|\mathbf{y}\right)$, and $p\left(^2\theta_3|\mathbf{y}\right)$ envelop their true values, and the estimation results are sufficiently accurate.
In the case of $l=8\,\mathrm{m}$ (\cref{fig:Pairplots2D}(B)), which corresponds to the correlation length of the true hydraulic conductivity field, 95\% HPDs for $p\left(^2\theta_1|\mathbf{y}\right)$, $p\left(^2\theta_2|\mathbf{y}\right)$, and $p\left(^2\theta_3|\mathbf{y}\right)$ envelop their true values, and these CIs are smaller than those seen in (\cref{fig:Pairplots2D}(A)). Furthermore, their means are nearly identical to the true values.
Additionally, in the bivariate marginal distributions, the 95\% HPD regions for $l=8\,\mathrm{m}$ are narrower than those for $l=4\,\mathrm{m}$. Overall, unlike the spatial-field-only estimation method, the proposed simultaneous estimation method provides more reliable quantification of uncertainties associated with the circular cavity boundary and the hydraulic conductivity spatial field.

\section{Concluding Remarks}
This paper proposes an inversion procedure for simultaneous geometry and spatial field estimation using the domain independence property of the K-L expansion. 
The method addresses the computational bottleneck in traditional simultaneous estimation methods caused by the repeated evaluation of the domain-dependent integral eigenvalue problem (IEVP) on evolving physical domains.
By solving the IEVP once on a fixed bounding domain and using these results throughout the inversion process, the proposed method avoids repeated costly computations, significantly improving computational efficiency over previous simultaneous estimation methods \cite{koch2021}. In particular, the expensive computations of the Moore-Penrose inverse to obtain the shape derivatives are completely eliminated.

Performance comparisons between the proposed method and the method of Koch et al. (2021) \cite{koch2021} reveal substantial computational gains. Two orders of magnitude improvement in the K-L expansion gradient computation time is observed with time complexity in the proposed method established as $O(n_\mathrm{e})$.  
For the 1D seepage flow problem, the time complexity for the computation of one leapfrog step (which includes the cost of the finite element forward solver and the derivative of the K-L expansion), in the proposed method, is also observed to be $O(n_\mathrm{e})$. This cost is similar to that incurred in the computation of the forward solver in 1D. In general, in higher dimensions, the time complexity of the forward solver lies in a range between $O(n_\mathrm{e})$ and $O(n_\mathrm{e}^3)$, implying that the $O(n_\mathrm{e})$ time complexity of derivative of the K-L expansion is no longer a bottleneck.
Additionally, for an approximate implementation of the domain independence property with a K-L expansion truncated to $M_1$ terms such that $\lambda_{M_1}>10^{-3}$, the estimation results of the proposed simultaneous estimation method are almost identical to those of the simultaneous estimation method of Koch et al. (2021) \cite{koch2021}. 
This confirms that the approximation made by using a truncated set of eigenpairs, computed on the bounding domain, in the K-L expansion is sufficient to represent the spatial random field accurately on updated domains during inversion. 

Methodological advantages of the simultaneous estimation method are also demonstrated. 
In the inverse problem for 1D seepage flow, geometry parameters are incorporated to define the unknown interface depths of a thin clay seam sandwiched between two homogeneous sandy layers. This approach enables the capture of abrupt jumps in posterior samples of the hydraulic conductivity spatial random field. In contrast, the spatial-field-only estimation method cannot detect such jumps, regardless of the size of the correlation length. Specifically, a large correlation length produces a smooth spatial distribution, whereas a small correlation length exhibits a large variance. In simultaneous estimation, although the 95\% CIs completely envelop the true hydraulic conductivity profile for all choices of correlation lengths, the width of the 95\% CIs varies with correlation lengths. This reveals the importance of selecting an appropriate correlation length.

The inverse problem for 2D seepage flow is set up in a domain containing an impermeable circular cavity of unknown location and size. 
The incorporation of geometry parameters defining the impermeable circular boundary allows for the implementation of accurate zero-flux BCs. As in the 1D case, smooth realizations from the posterior make it difficult to infer the uncertainty in the location of the circular cavity boundary using the spatial-field-only estimation method.
On the other hand, the proposed simultaneous estimation method yields accurate estimates of the hydraulic conductivity spatial field by providing straightforward estimates of the impermeable cavity boundary location and its uncertainties, thereby outperforming the spatial-field-only approach.
Additionally, it is observed that the incorporation of the geometry parameters can reduce the dimensionality of the inverse problem compared to the spatial-field-only approach. This is because the spatial discontinuities are parameterized by geometry parameters apriori, and the choice of a small correlation length is unnecessary.

Based on the numerical results, it is clear that an appropriate choice of the prior correlation length parameter is essential for obtaining accurate posterior estimates. One potential future extension to this study is the inference of this parameter during the Bayesian updating process. This poses the challenge that the formulation of the IEVP changes with changes in the correlation length. In \cite{latz2019}, a parametrized K-L expansion is proposed that could be implemented in this context. Another possible future research theme could be to combine this approach with an immersed boundary method, such as the finite cell method \cite{parvizian2007finite}, which would avoid remeshing. Finally, HMC faces challenges in identifying multi-modal posteriors, that often appear in practical engineering problems. The proposed method could implement HMC within a sequential Monte Carlo approach that enables efficient sampling of multimodal targets.

\begin{comment}
\section*{Declaration of generative AI and AI-assisted technologies in the writing process}
During the preparation of this work, the authors used ChatGPT, editGPT, and DeepL in order to proofread the text. After using these tools, the authors reviewed and edited the content as needed and take full responsibility for the content of the published article.
\end{comment}

\section*{Acknowledgements}
This work was supported by JSPS KAKENHI Grant Number JP22K18352.

\appendix
\section{Discrete Karhunen-Loève expansion and its derivative in HMC}\label{app:discreteKL}

The K-L expansion for a random vector is known as the discrete K-L expansion \cite{dur1998}. 
Let the random vector $\mathbf{X}\left(\omega\right)=\left[X\left(\mathbf{z}_{1},\omega\right),\ldots,X\left(\mathbf{z}_{n_{{\rm e}}},\omega\right)\right]^{\top}$ be the spatial discretization of $X\left(\mathbf{z},\omega\right)$, 
on a set of $n_{{\rm e}}$ discrete points $\left\{ \mathbf{z}_{i}\right\} _{i=1}^{n_{{\rm e}}}\subset\mathcal{D}$. In the discrete K-L expansion,
the following eigenvalue problem for the covariance matrix is solved:

\begin{equation}
\mathbf{C_{\mathcal{D}}}\mathbf{\Phi}_{i}=\lambda_{i}\mathbf{\Phi}_{i},\label{eq:EVP}
\end{equation}
where the covariance matrix $\mathbf{C}_{\mathcal{D}}\in\mathbb{R}^{n_{{\rm e}}\times n_{{\rm e}}}$ is a symmetric positive semi-definite matrix with elements $c_{ij}=C\left(\mathbf{z}_{i},\mathbf{z}_{j}\right)$,
$\lambda_{k}$ and $\mathbf{\Phi}_{k}\in\mathbb{R}^{n_{{\rm e}}}$ are eigenvalues and eigenvectors of $\mathbf{C}_{\mathcal{D}}$, and $\mathbf{\Phi}_{k}$ are orthogonal, that is, $\mathbf{\Phi}_{i}^{\top}\mathbf{\Phi}_{j}=\delta_{ij}$.
The eigen decomposition $\mathbf{C_{\mathcal{D}}}=\sum_{i=1}^{n_{{\rm e}}}\lambda_{i}\mathbf{\Phi}_{i}\mathbf{\Phi}_{i}^{\top}$
corresponds to Mercer's theorem for $C\left(\mathbf{z},\mathbf{z}^{*}\right)$.
The discrete K-L expansion is represented as

\begin{equation}
\mathbf{X}\left(\omega\right)=\overline{\mathbf{X}}+\sum_{i=1}^{n_{{\rm e}}}\sqrt{\lambda_{i}}\mathbf{\Phi}_{i}{}^{1}\theta_{i}\left(\omega\right),\label{eq:KL2}
\end{equation}
where $^{1}\theta_{i}\left(\omega\right)=
\frac{1}{\sqrt{\lambda_{i}}}
\left(\mathbf{X}\left(\omega\right)-\overline{\mathbf{X}}\right)^{\top}\mathbf{\Phi}_{i}$,
and $^{1}\theta_{i}$ satisfies $E\left[^{1}\theta_{i}\right]=0$, $ E\left[^{1}\theta_{i}{}^{1}\theta_{j}\right]=\delta_{ij}$.
The truncated discrete K-L expansion is then expressed as

\begin{equation}
\hat{\mathbf{X}}\left(\omega\right)=\overline{\mathbf{X}}+\sum_{i=1}^{M}\sqrt{\lambda_{i}}\mathbf{\Phi}_{i}{}^{1}\theta_{i}\left(\omega\right).\label{eq:truncated_KL2}
\end{equation}
The derivative of $\hat{\mathbf{X}}$ with respect to the K-L expansion
parameters $^{1}\bm{\uptheta}$ is given by

\begin{equation}
\frac{\partial\mathbf{\hat{\mathbf{X}}}}{\partial{}^{1}\theta_{i}}=\sqrt{\lambda_{i}}\bm{\Phi}_{i},
\end{equation}
and the derivative of $\hat{\mathbf{X}}$ with respect to the geometry parameters $^{2}\bm{\uptheta}$ is given by

\begin{equation}
\frac{\partial\mathbf{\hat{\mathbf{X}}}}{\partial{}^{2}\bm{\uptheta}}=\sum_{i=1}^{M}\left(\frac{1}{2\sqrt{\lambda_{i}}}\frac{\partial\lambda_{i}}{\partial{}^{2}\bm{\uptheta}}\bm{\Phi}_{i}+\sqrt{\lambda_{i}}\frac{\partial\bm{\Phi}_{i}}{\partial{}^{2}\bm{\uptheta}}\right){}^{1}\theta_{i}.
\end{equation}
The shape derivatives of eigenvalues and eigenvectors are analytically computed by the method of Magnus \cite{magnus1985}, and are given by

\begin{equation}
\frac{\partial\lambda_{i}}{\partial{}^{2}\bm{\uptheta}}=\bm{\Phi}_{i}^{\top}\frac{\partial\mathbf{C_{\mathcal{D}}}}{\partial{}^{2}\bm{\uptheta}}\bm{\Phi}_{i},
\end{equation}
\begin{equation}\label{eq:derivative_discreteKL}
\frac{\partial\bm{\Phi}_{i}}{\partial{}^{2}\bm{\uptheta}}=\left(\lambda_{i}\mathbf{I}-\mathbf{C_{\mathcal{D}}}\right)^{\dag}\frac{\partial\mathbf{C}_{\mathcal{D}}}{\partial{}^{2}\bm{\uptheta}}\bm{\Phi}_{i}.
\end{equation}

These derivatives are computed every time the parameters ${}^{2}\bm{\uptheta}$ are updated in HMC. In particular, the calculation of $\frac{\partial\bm{\Phi}_{i}}{\partial{}^{2}\bm{\uptheta}}$ requires the Moore-Penrose pseudoinverse $\left(\cdot\right)^{\dag}$ and is the main reason for the high computational cost in a naive implementation of simultaneous spatial-field and geometry estimation.

%\appendix
%\section{Example Appendix Section}
%\label{app1}


%%       \cite{<label>} ==> [1]

%%
%Example citation, See %\cite{lamport94}.

\bibliographystyle{elsarticle-num} 
\bibliography{reference}



%% \begin{thebibliography}{00}
%% 
%% %% For numbered reference style
%% %% \bibitem{label}
%% %% Text of bibliographic item
%% 
%% \bibitem{lamport94}
%%   Leslie Lamport,
%%   \textit{\LaTeX: a document preparation system},
%%   Addison Wesley, Massachusetts,
%%   2nd edition,
%%   1994.
%% 
%% \end{thebibliography}
\end{document}